\documentclass[12pt]{article}

\usepackage[francais,english]{babel}
\usepackage{amssymb}
\usepackage{amsmath,amsfonts}

\usepackage{tikz}
\usetikzlibrary{decorations.pathreplacing}
\usetikzlibrary{decorations.markings}
\usetikzlibrary{patterns}

\tikzset{marking1/.style = {decoration = {markings, mark = at position .5 with {\arrow[line width = 1pt]{>}}},postaction=decorate }}
\usepackage{color}

\colorlet{lightblue}{blue!40!}
\colorlet{lightyellow}{yellow!40!}
\colorlet{lightgreen}{green!40!}

\newcommand{\atl}[1]{\mathsf{aTL}_{#1}}
\newcommand{\tl}[1]{\mathsf{TL}_{#1}}
\newcommand{\q}{\mathfrak{q}}
\newcommand{\Q}{\tilde{\mathfrak{q}}}
\newcommand{\s}{\mathfrak{s}}

\begin{document}

\begin{center}

\Large{Topological defects in periodic RSOS models\\ and anyonic chains}

\vskip 1cm

\large{J. Bellet\^ete\,$^1$, A.M. Gainutdinov\,$^2$, J.L. Jacobsen\,$^{1,3,4}$, H. Saleur\,$^{1,5}$, T.S. Tavares\,$^1$}

\vspace{1.0cm}

{\sl\small $^1$  Institut de Physique Th\'eorique, Universit\'e Paris Saclay, CEA, CNRS, 91191 Gif-sur-Yvette, France\\}

{\sl\small $^2$
Institut Denis Poisson, CNRS, Universit\'e de Tours, Universit\'e d'Orl\'eans, \\Parc de Grammont, 37200 Tours, France\\}

{\sl\small $^3$
Laboratoire de Physique de l'\'Ecole Normale Sup\'erieure, ENS,\\ Universit\'e PSL,CNRS, Sorbonne Universit\'e, \\Universit\'e de Paris, F-75005 Paris, France\\}

{\sl\small $^4$
Sorbonne Universit\'e, \'Ecole Normale Sup\'erieure, CNRS, \\
 Laboratoire de Physique Th\'eorique (LPT ENS), 75005 Paris, France \\}

{\sl\small $^5$
Department of Physics,
  University of Southern California, Los Angeles, CA 90089-0484,
   USA \\}

\end{center}


\begin{abstract}
We provide a  lattice regularization of  all topological defects in minimal models CFTs using RSOS and anyonic spin chains.

For defects of type $(1,s)$, we connect our result with   the ``topological symmetry'' initially identified in Fibonacci anyons \cite{Feiguinetal}, and the center of the affine Temperley-Lieb algebra discussed in \cite{BGJST}. We show that the topological nature of the defects is exact on the lattice as well. Our defects of type $(r,1)$, in contrast, are only topological in the continuum limit. Identifications are obtained by a mix of algebraic and Bethe-ansatz techniques. Most of our discussion is framed in a Hamiltonian (or transfer matrix) formalism, and direct and crossed channel are both discussed in detail. For defects of type $(1,s)$, we also show how to implement their fusion, which turns out to reproduce the tensor product of the underlying monoidal category used to build the anyonic chain.

\end{abstract}


\newpage

\tableofcontents

\newpage
\section{Introduction}

A defect in a two-dimensional CFT is a line of inhomogeneity along which certain matching conditions for the fields are implemented. Defects come in various types. Generically, they are not scale invariant, and give rise to properties that flow under the renormalization group (RG). Fixed points of the RG correspond to conformal defects, a special class of which is called \emph{topological}.

To define topological defects in a CFT more precisely, it is useful to consider radial quantization and imagine  that the defect line runs along the unit circle. Equivalently, on can consider the geometry of a  cylinder, with the defect line a non-contractible cycle in the periodic (imaginary time) direction. In this formulation, the defect becomes an operator $D$ acting on the Hilbert space of the theory. Such defect is topological iff $D$ commutes with the left and right Virasoro algebras:
\begin{equation}
[D,L_n]=[D,\bar{L}_m]=0\label{VirComm},
\end{equation}
which corresponds to having independently $T,\bar{T}$ continuous across the defect line. In contrast, the defect is conformal if only it commutes 
 with a diagonal sub-algebra
\begin{equation}
[D,L_n-\bar{L}_{-n}]=0,
 \end{equation}
 which only corresponds to having $T-\bar{T}$ continuous across the  defect line.

When the defect is topological, the defect line can be deformed continuously without changing the partition function, which justifies its name.

 The construction and classification of topological defects in CFT has been an active topic for many years - see e.g. \cite{Fr1,Fr2,FRS,Fu,PZ,PZ1} for early references - with many applications ranging from condensed matter (see e.g. \cite{KL}) to string theory (see e.g. \cite{BBDO,Ba,BB,BG}). A particularly interesting aspect of the topic concerns the relationship  between the  topological defects  in the CFTs and their possible lattice regularizations - a question whose origin goes back  to the pioneering work in \cite{KC}.  It has  recently been suggested \cite{AMF} that these defects could be built directly at the level of the lattice, and their algebra studied in detail without ever taking the continuum limit - hence providing, in effect, a painless way to identify CFTs and critical exponents. While we do not believe this program is viable in general, the problem of how to build lattice versions of topological defects is nevertheless interesting for several reasons. On the one hand, it may provide a handle to understand ill-identified bulk CFTs, such as the logarithmic CFTs that appear in the description of geometrical problems (polymers, percolation) or phase transitions in disordered systems (like the plateau transition in the integer quantum Hall effect). On the other hand, it can provide precious insight even on well known CFTs, in particular by giving access to potentially hidden symmetries and or allowing the use of concepts such as ``topological protection'' in the analysis of specific models.

 We will mostly restrict in this paper to the simplest of all cases, the diagonal, unitary, minimal models. Parametrizing the central charge with
 \begin{equation}
 c=1-{6\over p(p+1)},
 \end{equation}
 it is known that these models admit the modular invariant partition function \cite{CIZ87} 
 \begin{equation}
 Z={1\over 2}\sum_{r=1}^{p-1}\sum_{s=1}^p |\chi_{rs}|^2,\label{CFTz}
 \end{equation}
 where the $\chi_{rs}$ are the characters of the irreducible Virasoro algebra representation with conformal weight given by the Kac formula
 \begin{equation}
 h_{rs}={((p+1)r-ps)^2-1\over 4p(p+1)}.
 \end{equation}
 Topological defects for these models have been classified \cite{PZ,PZ1}. They turn out to be in one to one correspondence with the primary fields of the bulk theory, and thus parametrized by a single pair of Kac labels, $(r,s)$. We denote the chiral and anti-chiral components of these fields by $\phi_{rs}$ and $\bar{\phi}_{rs}$ respectively.

 The problem of building lattice versions of these defects - that is, defect lines expressed in terms of lattice variables, that go over to the corresponding topological defect in the continuum limit - has been tackled in various ways. For the  RSOS lattice version of the CFTs  (\ref{CFTz}), a general  construction was proposed in \cite{CMOP,CMOP1} based on a rather complex application of integrability techniques. In the equivalent anyonic formulation, defects of type $(1,s)$ were identified early, and connected with the existence of ``topological symmetries'' \cite{Gilsetal,Feiguinetal,BuicanGromov}. A wealth of alternative constructions has been proposed for specific models in the series (\ref{CFTz}) (such as Ising and tricritical Ising) \cite{Grimm,AMF,LVV}.

 Our interest in the topic stems from our study of logarithmic CFTs, and the question of topological defects in models such as percolation \cite{sl21}. To make progress in this direction, it is crucial to have a better understanding of the construction of defects at the level of the lattice models. This can be achieved, we believe, by following an algebraic route.

In a previous paper \cite{BGJST},  we proposed a general construction of a certain type of topological defects based on new results about the center of the affine Temperley-Lieb algebra. While this construction is aimed at the study of defects in general models (like the $Q$-state Potts model, percolation, etc.), it should also apply to the specific case of RSOS models - or, equivalently, anyonic spin chains. Our purpose in the present note is  to discuss this application, and thus, it turns out, re-visit the construction of defects of type $(1,s)$ in anyonic models. We will then propose a simple generalization to construct also defects of type $(r,1)$, and then, finally, the most general $(r,s)$ defects.

Since we wish to mostly discuss spin chains here, we shall naturally use a Hamiltonian formulation of the 2D CFT, or, more generally, a transfer matrix formulation, with the lattice model defined on a cylinder, which can be, if necessary, closed into a torus. In this case,
there are obviously two possible types of defect lines. Those running parallel to the axis of the cylinder (and thus, associated with a ``defect hamiltonian'', and those running perpendicular to that axis (and thus, associated with a ``defect operator''). We shall refer to these two situations as direct and crossed channel respectively. Note that, after closing the cylinder into a torus, these two choices
are easily related within the CFT using modular transformations. From the point of view of the lattice models however, such relation is more complicated, albeit still possible is some cases \cite{PS,nextpaper}.

Section \ref{sec:anyons.def} contains the basic definitions of the models under consideration in this work as well as the notation we will be using. A generic anyon chain is first introduced and the periodic $ \mathsf{A}_{p}$ RSOS model is then presented. While it has been known for a long time that these two models are intimately related, we could not find a general map linking the two; we present a simple one described through the action of the Temperley-Lieb algebra, chosen to make the calculation of defect properties particularly simple.

Section \ref{sec:topdefect.introduction} introduces our construction of lattice topological defects, which falls into three large classes which we label as chiral, anti-chiral, and inter-chiral. The first two classes, grouped into what we call $s$-\emph{type}, are central elements of the affine Temperley-Lieb algebra, while the third class, which we call $r$-\emph{type}, are not. We also explain the topological invariance, or non-invariance, of our construction.

Section \ref{sec:topdefect.merging} explains how the defects of $s$-type can be \emph{merged} or \emph{fused} together; in essence this section explains how to relate models carrying multiple defects to those carrying only one. In particular, we show how the fusion of lattice defect operators reproduce the product rules of the underlying fusion category in both channels. Merging of defects in mixed channels - some in the direct channel with others in the crossed channel, is also discussed.

Section \ref{sec:topdefect.analytics} present various examples of defects in three specific cases: the $A_{3}$ model, related to the Ising model, the $A_{4}$ model which is related to the Fibonacci anyon chain, as well as the $A_{5}$ model related to the three-state Potts models. Most of these results were obtained analytically or numerically, as the action of the $r$-type defects seemingly cannot be studied using purely algebraic methods, like those of $s$-type. We give a brief overview of the methods we used to gather these results.

\section{Anyonic chains}\label{sec:anyons.def}
We introduce here our formulation of anyonic chains built from the $\mathsf{A}_{p}$ strict fusion category $\mathcal{C}_{p} $, i.e. the simple objects are $\mathsf{A}_{p} = \lbrace 1,2,\hdots, p\rbrace $ with tensor product
	\begin{equation*}
		r \otimes s = \sum_{\underset{\text{step } = 2}{k= 1+ |r-s|}}^{\mathsf{min}(r+s - 1, 2p - (r+s) + 1)} k,
	\end{equation*}
the unitors and associators are all identities, and where every element is self-dual (both left and right). Our conventions for the $F$-symbols are given in appendix \ref{sec:Fsymbols}; they were taken from \cite{Fsymbols} (equation 110), except that their $\q$-numbers should be taken as $\lfloor x \rfloor_{\q} = (-1)^{x} \frac{\q^{x} - \q^{-x}}{\q - \q^{-1}}$.

\subsection{The Hilbert space}
The Hilbert space of the chain $\mathcal{H}_{a}[n]$, for $a \in \mathsf{A}_{p}$, is defined as
\begin{equation}\label{eq:anyonHamil}
	\mathcal{H}_{a}[n] \equiv \bigoplus_{x_{0} \in \mathsf{A}_{p}}\mathsf{Hom}_{\mathcal{C}_{p}}(x_{0} \otimes a^{\otimes n},x_{0}).
\end{equation}
A basis of this vector space can be constructed from \emph{fusion trees}:
\begin{equation}
	|x_{0}, x_{1}, \hdots, x_{n-1},x_{0} \rangle \equiv \;
	\begin{tikzpicture}[baseline = {(current bounding box.center)},scale = 2/3]
	\node[anchor = east] at (0,0) {$\footnotesize{x_{0} }$};
	\node[anchor = south] at (1,1) {$\footnotesize{a}$};
	\node[anchor = south] at (2,1) {$\footnotesize{a}$};
	\node[anchor = south] at (3,1) {$\footnotesize{\hdots}$};
	\node[anchor = south] at (4,1) {$\footnotesize{a}$};
	\node[anchor = south] at (5,1) {$\footnotesize{a}$};
	\node[anchor = west] at (6,0) {$\footnotesize{x_{0}}$};
	\node[anchor = north] at (3/2,0) {$\footnotesize{x_{1}}$};
	\node[anchor = north] at (3,0) {$\footnotesize{\hdots}$};
	\node[anchor = north] at (9/2,0) {$\footnotesize{x_{n-1}}$};
	\draw[black, line width = 1pt] (0,0) -- (6,0);
	\foreach \r in {1,2,4,5}{
		\draw[black, line width = 1pt] (\r,0) -- (\r,1);
	}
	\end{tikzpicture}\;,
\end{equation}
which are graphical representations of a sequence of canonical projections from $x_{i}\otimes a \to x_{i+1} $; these trees therefore represents morphisms
\begin{equation*}
|x_{0},x_{1},\hdots \rangle:( \hdots((x_{0}\otimes a) \otimes a)\hdots)\otimes a ) \to x_{0}.
 \end{equation*}
 Different choice of basis could be obtained by simply changing the order of the parentheses. One can define a Hamiltonian acting on this Hilbert space as
\begin{equation}
	H_{a}[n] \equiv \sum_{i=1}^{n} X^{a}_{i}[n],\label{Hamil}
\end{equation}
where
\begin{equation}
	X^{a}_{i}[n] \equiv \;
	\begin{tikzpicture}[baseline = {(current bounding box.center)}, every node/.style={scale = 2/3}, scale = 1/2]
	\foreach \r in {1,3,6,8}{
	\draw[black, line width = 1pt] (\r,0) -- (\r,3);
	\node[anchor = south] at (\r,3) {$\footnotesize{a}$};
	\node[anchor = north] at (\r,0) {$\footnotesize{a}$};
	}
	\node[anchor = south] at (2,1) {$\footnotesize{\hdots}$};
	\node[anchor = south] at (7,1) {$\footnotesize{\hdots}$};
	\node[anchor = south] at (4,3) {$\footnotesize{a}$};
	\node[anchor = north] at (4,0) {$\footnotesize{a}$};
	\node[anchor = south] at (5,3) {$\footnotesize{a}$};
	\node[anchor = north] at (5,0) {$\footnotesize{a}$};
	\draw[black, line width = 1pt] (4,0) -- (4.5,1) -- (5,0) -- (4.5,1) -- (4.5,2) -- (4,3) -- (4.5,2) -- (5,3);
	\node[anchor = west] at (4.5,1.5) {$\footnotesize{1}$};
	\draw[decorate, decoration = {brace, mirror, amplitude = 3 pt}, yshift = -3pt] (.5,-.5) -- (3.5,-.5) node [midway,yshift = -10pt] {\footnotesize{i-1}};
	\draw[decorate, decoration = {brace, mirror, amplitude = 3 pt}, yshift = -3pt] (5.5,-.5) -- (8.5,-.5) node [midway,yshift = -10pt] {\footnotesize{n-i-1}};
	\end{tikzpicture} \; \in \mathsf{End}_{\mathcal{C}_{p}}(a^{\otimes n}) ,\qquad i \neq n,
\end{equation}
\begin{equation}
	X^{a}_{n}[n] \equiv \;
	\begin{tikzpicture}[baseline = {(current bounding box.center)}, every node/.style={scale = 2/3}, scale = 1/2]
	\draw[black, line width = 1pt] (0,0) -- (-.5,1) -- (-1,0) -- (-.5,1) -- (-.5,2) -- (0,3) -- (-.5,2) -- (-1,3);
	\filldraw[white] (-1.5,0) -- (-.75,0) -- (-.75,3) -- (-1.5,3) -- (-1.5,0);
	\draw[black, line width = 1pt, dashed] (-.75,0) -- (-.75,3);
	\foreach \r in {1,2,4,5}{
	\draw[black, line width = 1pt] (\r,0) -- (\r,3);
	\node[anchor = south] at (\r,3) {$\footnotesize{a}$};
	\node[anchor = north] at (\r,0) {$\footnotesize{a}$};
	}
	\node[anchor = south] at (3,1) {$\footnotesize{\hdots}$};
	\node[anchor = south] at (0,3) {$\footnotesize{a}$};
	\node[anchor = north] at (0,0) {$\footnotesize{a}$};
	\node[anchor = south] at (6,3) {$\footnotesize{a}$};
	\node[anchor = north] at (6,0) {$\footnotesize{a}$};
	\node[anchor = west] at (-.5,1.5) {$\footnotesize{1}$};
	\draw[black, line width = 1pt] (6,0) -- (6.5,1) -- (6.5,2) -- (6,3);
	\filldraw[white] (-1.5,0) -- (6.25,0) -- (6.25,3) -- (7.5,3) -- (6.25,0);
	\draw[black, line width = 1pt, dashed] (6.25,0) -- (6.25,3);
	\draw[decorate, decoration = {brace, mirror, amplitude = 3 pt}, yshift = -3pt] (.5,-.5) -- (5.5,-.5) node [midway,yshift = -10pt] {\footnotesize{n-2}};
	\end{tikzpicture} \; \in \mathsf{End}_{\mathcal{C}_{p}}(a^{\otimes n}) \; ,
\end{equation}
where one understand that the two opposing dashed lines are identified, so that $X^{a}_{n}[n]$ connects the first and the last tensorand in $a^{\otimes n}$. These operators act on the fusion trees from the top, i.e.
\begin{align*}
	X^{a}_{1}[2]|x_{0},x_{1} \rangle & \equiv \;
	\begin{tikzpicture}[baseline = {(current bounding box.center)}, every node/.style={scale = 2/3}, scale = 1/2]
	\draw[black, line width = 1pt] (0,0) -- (3,0);
	\node (a1) at (1,1) {$\footnotesize{a}$};
	\node (a2) at (2,1) {$\footnotesize{a}$};
	\node (a3) at (1,4) {$\footnotesize{a}$};
	\node (a4) at (2,4) {$\footnotesize{a}$};
	\draw[black, line width = 1pt] (a1) -- (1.5,2) -- (a2);
	\draw[black, line width = 1pt] (a3) -- (1.5,3) -- (a4);
	\draw[black, line width = 1pt] (1.5,2) -- (1.5,3);
	\draw[black, line width = 1pt] (1,0) -- (a1);
	\draw[black, line width = 1pt] (2,0) -- (a2);
	\node[anchor = west] at (1.5,2.5) {$\footnotesize{1}$};
	\node[anchor = north ] at (.5,0) {$\footnotesize{x_{0}}$};
	\node[anchor = north ] at (1.5,0) {$\footnotesize{x_{1}}$};
	\node[anchor = north ] at (2.5,0) {$\footnotesize{x_{0}}$};
	\end{tikzpicture} \; = \;
	\sum_{y \in \mathsf{A}_{p}} (\mathsf{F}_{x_{0},a,a}^{x_{0}})_{x_{1}}^{y} \;
	\begin{tikzpicture}[baseline = {(current bounding box.center)}, every node/.style={scale = 2/3}, scale = 1/2]
	\draw[black, line width = 1pt] (0,-1) -- (3,-1);
	\node (a1) at (1,1) {$\footnotesize{a}$};
	\node (a2) at (2,1) {$\footnotesize{a}$};
	\node (a3) at (1,4) {$\footnotesize{a}$};
	\node (a4) at (2,4) {$\footnotesize{a}$};
	\draw[black, line width = 1pt] (a1) -- (1.5,2) -- (a2);
	\draw[black, line width = 1pt] (a3) -- (1.5,3) -- (a4);
	\draw[black, line width = 1pt] (1.5,2) -- (1.5,3);
	\draw[black, line width = 1pt] (1.5,-1) -- (1.5,0) -- (a1);
	\draw[black, line width = 1pt] (1.5,-1) -- (1.5,0) -- (a2);
	\node[anchor = west] at (1.5,2.5) {$\footnotesize{1}$};
	\node[anchor = north ] at (.5,-1) {$\footnotesize{x_{0}}$};
	\node[anchor = west] at (1.5,-.5) {$\footnotesize{y}$};
	\node[anchor = north ] at (2.5,-1) {$\footnotesize{x_{0}}$};
	\end{tikzpicture} \; = \; \frac{(\mathsf{F}_{x_{0},a,a}^{x_{0}})_{x_{1}}^{1}}{(\mathsf{F}_{a,a,a}^{a})_{1}^{1}} \;
	\begin{tikzpicture}[baseline = {(current bounding box.center)}, every node/.style={scale = 2/3}, scale = 1/2]
	\draw[black, line width = 1pt] (0,-1) -- (3,-1);
	\node (a1) at (1,1) {$\footnotesize{a}$};
	\node (a2) at (2,1) {$\footnotesize{a}$};
	\draw[black, line width = 1pt] (1.5,-1) -- (1.5,0) -- (a1);
	\draw[black, line width = 1pt] (1.5,-1) -- (1.5,0) -- (a2);
	\node[anchor = north ] at (.5,-1) {$\footnotesize{x_{0}}$};
	\node[anchor = west] at (1.5,-.5) {$\footnotesize{1}$};
	\node[anchor = north ] at (2.5,-1) {$\footnotesize{x_{0}}$};
	\end{tikzpicture} \\
	& = \frac{(\mathsf{F}_{x_{0},a,a}^{x_{0}})_{x_{1}}^{1}}{(\mathsf{F}_{a,a,a}^{a})_{1}^{1}} \sum_{y \in \mathsf{A_{p}}} (\mathsf{F}_{x_{0},a,a}^{x_{0}})_{y}^{1} \;
	\begin{tikzpicture}[baseline = {(current bounding box.center)}, every node/.style={scale = 2/3}, scale = 1/2]
	\draw[black, line width = 1pt] (0,0) -- (3,0);
	\node (a1) at (1,1) {$\footnotesize{a}$};
	\node (a2) at (2,1) {$\footnotesize{a}$};
	\draw[black, line width = 1pt] (1,0) -- (a1);
	\draw[black, line width = 1pt] (2,0) -- (a2);
	\node[anchor = north ] at (.5,0) {$\footnotesize{x_{0}}$};
	\node[anchor = north ] at (1.5,0) {$\footnotesize{y}$};
	\node[anchor = north ] at (2.5,0) {$\footnotesize{x_{0}}$};
	\end{tikzpicture} \;.
\end{align*}
 Using the conventions of Appendix \ref{sec:Fsymbols}, one shows that (see Appendix \ref{app:TL-relations} for a proof of the last relation):
\begin{align}
	X^{a}_{i}[n]X^{a}_{i}[n] & = \frac{1}{(\mathsf{F}_{a,a,a}^{a})_{1}^{1}} X^{a}_{i}[n], \notag\\
	X^{a}_{i}[n]X^{a}_{j}[n] & = X^{a}_{j}[n]X^{a}_{i}[n], \qquad |i-j| \geq 2, \label{TL-relations} \\
	X^{a}_{i}[n]X^{a}_{i \pm 1}[n]X_{i}^{a}[n] &= (\mathsf{F}_{1,a,a}^{1})_{a}^{1} X_{i}^{a}[n], \notag
\end{align}
where it is understood that the index of these operators are periodic ($X^{a}_{0}[n]\equiv X^{a}_{n}[n] $, $X^{a}_{n+1}[n]\equiv X^{a}_{1}[n] $). It follows that these operators generate a representation of the periodic Temperley-Lieb algebra on $n$ sites, provided that $(\mathsf{F}_{1,a,a}^{1})_{a}^{1} \neq 0$, by defining
\begin{equation}
	E^{a}_{i}[n] \equiv ((\mathsf{F}_{1,a,a}^{1})_{a}^{1})^{-1/2}X^{a}_{i}[n], \qquad \rightarrow E^{a}_{i}[n]E^{a}_{i}[n] = \frac{((\mathsf{F}_{1,a,a}^{1})_{a}^{1})^{1/2}}{(\mathsf{F}_{a,a,a}^{a})_{1}^{1}}E^{a}_{i}[n].
\end{equation}
With our convention (appendix \ref{sec:Fsymbols}), we find
\begin{equation}
	(\mathsf{F}_{1,a,a}^{1})_{a}^{1} = 1 , \qquad (\mathsf{F}_{a,a,a}^{a})_{1}^{1} = \frac{\q - \q^{-1}}{\q^{a} - \q^{-a}},
\end{equation}
with $\q = e^{i \frac{\pi}{p+1}}$. It follows that these anyon chains can be interpreted as loop models with loop weight $\beta = \frac{\q^{a} - \q^{-a}}{\q - \q^{-1}}$.

\subsection{Anyonic chains and the RSOS models}\label{sec:RSOS}

\begin{figure}
\begin{center}
\begin{tikzpicture}[scale = 2/3, baseline = {(current bounding box.center)}]
	\foreach \r [count = \i ]in {3,4,3,2,1,2,3,4,5,4,5,4,3,2,1}{
		\coordinate (S\i) at (\i,\r);
	}
	\foreach \i [evaluate=\i as \x using \i+1] in {1,...,14}{
		\draw[black, line width = 2pt] (S\i) -- (S\x);
	};
\draw[black, line width = 2pt] (0,1) -- (0,5);
\foreach \r in {1,...,5}{
	\filldraw[black] (0,\r) circle (4 pt);
	\node[anchor = east] at (-.25,\r) {\small{\r}};
	};
\draw[black, line width = 2pt] (1,0) -- (15,0);
\foreach \r in {0,...,14}{
	\draw[black, dashed, line width = .5pt] (\r +1,1) -- (\r +1, 5);
	\filldraw[black] (\r + 1,0) circle (4 pt);
	\node[anchor = north] at (\r +1,-.25) {\small{\r}};
	};
\end{tikzpicture}
\end{center}
\caption{One RSOS path on $\mathsf{A}_{5}$ with $14$ steps, starting from $3$ and ending at $1$. The scale on the left shows the position on the the Dynkin diagram, while the scale at the bottom counts the steps.}\label{fig:M31Path}
\end{figure}
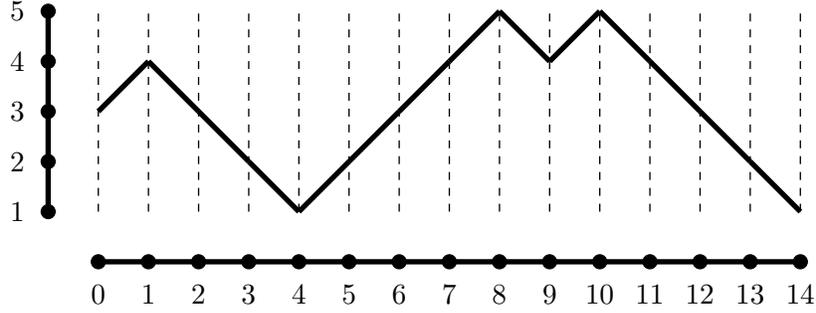

The simplest example of anyonic chains are the famous $\mathsf{A}_{p}$ RSOS models; these describe random single-step walks on the $A_{p}$ Dynkin diagram which start and end at the same point. For convenience we encode walks in vectors by their history, i.e. the vector $|x_{0}, x_{1}, x_{2}, \hdots, x_{n} \rangle $ represents the walk that starts at $x_{0}$, then takes a step to $x_{1}$, then a step to $x_{2}$, etc. Figure \ref{fig:M31Path} shows the path corresponding to $|3,4,3,2,1,2,3,4,5,4,5,4,3,2,1\rangle$ on $\mathsf{A}_{5}$.

The action of the $\atl{n}$ generators on these paths is given by \cite{PasquierRSOS}: 
\begin{align}
	e_{i}|x_{0},x_{1},\hdots, x_{n} \rangle = & \delta_{x_{i-1}, x_{i+1}} \frac{ [ x_{i}]_{\q}^{1/2} }{[x_{i-1}]_{\q}} \big( [x_{i-1}+1]_{\q}^{1/2}|x_{0},x_{1}, \hdots, x_{i-1}, x_{i-1}+1, x_{i+1}, \hdots, x_{n}\rangle \notag \\
	& + [x_{i-1}- 1]_{\q}^{1/2}|x_{0},x_{1}, \hdots, x_{i-1}, x_{i-1}-1, x_{i+1}, \hdots, x_{n}\rangle \big),
\end{align}
where we used the $\q$ -numbers $[k]_{\q} \equiv \frac{\q^{k} - \q^{-k}}{\q - \q^{-1}}$, and here $ \q = e^{i\frac{\pi}{p+1}} $. One can check directly that these satisfy the defining relations of the periodic Temperley-Lieb algebra, with loop weight $[2]_{\q}  =  \q + \q^{-1} $.The Hilbert space of this model is then simply the span of those paths which starts and end at the same point (any point) with the Hamiltonian
\begin{equation}\label{eq:RSOShamiltonian}
	H[n] = \sum_{i=1}^{n}e_{i}.
\end{equation}

Note that while the parallel with the anyon chain at $a= 2 $ seems obvious by simply inspecting the expression of the Temperley-Lieb generators, the exact correspondence between the two models is a bit more subtle. Indeed, the basis of the map between the two models is highly dependent on the specific convention chosen for the $F$-symbols, and this will in turn affect the matrix representing the generators $e_{i}$. However, it is known that the Hilbert space of the periodic RSOS model can be generated from $p$ initial states\footnote{It is actually possible to do it with a single initial state, but it is more convenient to take $p$ of them.}. One can thus identify the image of these states in the anyon chain, and use the algebra relations to obtain the image of the other ones. A very simple choice of generators is given by:
\begin{align}
	V_{1}[n] & \equiv \; |1,2,1,2,\hdots,2,1 \rangle,\\
	V_{r}[n] & \equiv \; \left(\frac{[r+1]_{\q}}{[r]_{\q}} \right)^{-n/2}\omega |r,r+1,r,r+1,\hdots,r+1,r\rangle, \quad r \neq 1,p, \\
	V_{p}[n] & \equiv \; |p,p-1,p,p-1,\hdots,p-1,p \rangle ,
\end{align}
where
\begin{equation}
 \omega \equiv (\q + \q^{-1})^{-n/2} e_{1}e_{3}\hdots e_{n-1}.
\end{equation}
The normalization of these vectors make the calculation of the topological charges easier (see \eqref{eq:hoopcalculation}). The corresponding generators in the anyon chain are:
\begin{align}\label{eq:basis.hilbert}
 U_{r}[n] & \equiv \;
 	\begin{tikzpicture}[baseline = {(current bounding box.center)}, every node/.style={scale = 1}, scale = 2/3]
	\foreach \r in {2,3,4,5,8,9}{
	\node (a\r) at (\r,3) {$\footnotesize{2}$};
	}
	\foreach \r in {5,9,17}{
	\draw[black, dashed, line width = 1pt] (\r/2,2) -- (\r/2,1);
	}
	\draw[black, line width = 1pt] (a2) -- (5/2,2);
	\draw[black, line width = 1pt] (a3) -- (5/2,2);
	\draw[black, line width = 1pt] (a4) -- (9/2,2);
	\draw[black, line width = 1pt] (a5) -- (9/2,2);
	\draw[black, line width = 1pt] (a8) -- (17/2,2);
	\draw[black, line width = 1pt] (a9) -- (17/2,2);
	\node (a0) at (1,1) {$\footnotesize{r}$};
	\node (a10) at (10,1) {$\footnotesize{r}$};
	\node[anchor = north] at (7/2,1) {$\footnotesize{r}$};
	\node[anchor = north] at (11/2,1) {$\footnotesize{r}$};
	\node[anchor = north] at (15/2,1) {$\footnotesize{r}$};
	\node[anchor = north] at (13/2,2) {$\hdots$};
	\draw[black, line width = 1pt] (a0) -- (a10);
	\end{tikzpicture} \;,
\end{align}
where the vertical dashed lines all carry the identity object, and the map between the two spaces is simply
\begin{equation}
	V_{r}[n] \longleftrightarrow U_{r}[n].
\end{equation}

\section{A proposal for topological defects}\label{sec:topdefect.introduction}

We discuss here the realization of topological defects as algebraic operations, or \emph{surgeries}, acting on these anyon chains. We first discuss the so-called \emph{direct channel}, that is when the defect is propagating parallel to the time evolution of the system, then describe the \emph{crossed channel}, where the defects are orthogonal to it. We do not prove at this stage that we are building topological defects indeed: this is discussed in later sections.  Furthermore, while we will present these operations one at a time, more complex defects can be constructed by simply performing multiple of them on the same chain; for instance, a defect of type $(r,s)$ is obtained by first inserting a defect of type $r$ and then a defect of type $s$, or vice versa.

\subsection{Defects in the direct channel}
In the direct channel, defects can be interpreted as a deformation affecting the Temperley-Lieb generators appearing in the Hamiltonian, i.e.
\begin{equation}\label{eq:hamilt.1defect.direct}
	H_{a}[n] = \sum_{i = 0}^{n-1} E^{a}_{i}[n]  \longrightarrow \tilde{H}^{(k)}_{a}[n] = \sum_{i = 0}^{n-2} \tilde{E}^{a}_{i}[n+k]  + \tilde{E}^{a,(k)}_{n-1}[n+k],
\end{equation}
where $k$ is the \emph{width} of the defect, and it is understood that $E^{a}_{0}[n] \equiv E^{a}_{n}[n]$; the deformed generators are, for $i\neq n-1 $,
\begin{equation}\label{def:vanilladefect}
	\tilde{E}^{a}_{i}[n+k]  = E^{a}_{i}[n+k] (1_{a^{\otimes n-1}} \otimes J_{k+1} \otimes 1_{a}),
\end{equation}
where the projector $J_{k+1}$ is a primitive idempotent of $\tl{k}$. There are usually many choices of idempotents producing equivalent defects; we focus here on the simplest choice, the \emph{Jones-Wenzl} projectors, which are only defined for $a=2$ and $k \leq p$:
\begin{equation*}
J_{k+1} \equiv 
	\frac{[k+1]^{1/2}_{\q}}{[2]^{k/2}_{\q}} \;
	\begin{tikzpicture}[baseline = {(current bounding box.center)},xscale = 2/3,yscale = 1/2]
	\draw[black, line width = 1pt] (1,3) -- (11/3,1/3);
	\draw[black, line width = 1pt] (2,3) -- (5/3,7/3);
	\draw[black, line width = 1pt] (3,3) -- (7/3,5/3);
	\draw[black, line width = 1pt] (4,3) -- (3,1);
	\draw[black, line width = 1pt] (5,3) -- (11/3,1/3);
	\draw[black, line width = 1pt] (11/3,1/3) -- (11/3,-2/3);
	\draw[black, line width = 1pt] (11/3,-2/3) -- (5,-10/3);
	\draw[black, line width = 1pt] (9/3,-4/3) -- (4,-10/3);
	\draw[black, line width = 1pt] (7/3,-6/3) -- (3,-10/3);
	\draw[black, line width = 1pt] (5/3,-8/3) -- (2,-10/3);
	\draw[black, line width = 1pt] (11/3,-2/3) -- (1,-10/3);
	\foreach \r in {1,2,3,4,5}{
	\node[anchor = south] at (\r,3) {\footnotesize{2}};
	\node[anchor = north] at (\r,-10/3) {\footnotesize{2}};
	}
	\node[anchor = north] at (5/3,7/3) {\footnotesize{3}};
	\node[anchor = north] at (7/3,5/3) {\footnotesize{4}};
	\node[anchor = south] at (7/3,-6/3) {\footnotesize{4}};
	\node[anchor = south] at (5/3,-8/3) {\footnotesize{3}};
	\foreach \r in {1,...,3}{
		\fill[black] (10/3 + \r/6 , 5/3 - \r/6) circle (1/16);
		\fill[black] (11/3 + \r/6 , -8/3 + \r/6) circle (1/16);
	}
	\node[anchor = east] at (11/3,-1/3) {\footnotesize{k+1}};
	\draw[decorate, decoration = {brace, amplitude = 3 pt}, yshift = 3pt] (1,4) -- (5,4) node [midway,yshift = 7pt] {\footnotesize{k}};
	\end{tikzpicture} \; \in \mathsf{End}_{\mathcal{C}_{p}}(2^{\otimes k}) \;,
\end{equation*}
with $J_{2}$ simply being the identity on $2$. The expression of $ \tilde{E}^{a,(k)}_{n-1}[n+k]$ is however a bit more subtle; we find three non-equivalent choices: two $s$-type defects, which we refer to as \emph{over} or \emph{under}, respectively, and one $r$-type defect.

\subsubsection{Defects of type $s$}
There are two choice of defects of type $s$; the \emph{over} case is
\begin{equation}
\tilde{E}^{2,(k)}_{n-1}[n+k]  \equiv \;
	\begin{tikzpicture}[baseline = {(current bounding box.center)},xscale = 1/2,yscale = 1/3]
	\foreach \s in {2,3,5,6}{
	\draw[black, line width = 1pt] (\s,1) -- (\s,9);
	}
	\draw[white, line width = 3pt] (1,1) -- (4,4);
	\draw[white, line width = 3pt] (4,4) -- (7,1);
	\draw[white, line width = 3pt] (1,9) -- (4,6);
	\draw[white, line width = 3pt] (4,6) -- (7,9);
	\draw[black, line width = 1pt] (1,9) -- (4,6);
	\draw[black, line width = 1pt] (4,6) -- (7,9);
	\draw[black, line width = 1pt] (1,1) -- (4,4);
	\draw[black, line width = 1pt] (4,4) -- (7,1);
	\draw[black, dashed, line width = 1pt] (4,4) -- (4,6);
	\foreach \r in {1,...,3}{
		\fill[black] (2 + \r/4 ,3/2 ) circle (1/16);
		\fill[black] (2 + \r/4 ,17/2 ) circle (1/16);
		\fill[black] (5 + \r/4 ,3/2 ) circle (1/16);
		\fill[black] (5 + \r/4 ,17/2 ) circle (1/16);
	}
	\draw[decorate, decoration = {brace, mirror, amplitude = 3 pt}, yshift = -3pt] (1.5,.5) -- (6.5,.5) node [midway,yshift = -7pt] {\footnotesize{k}};
	\end{tikzpicture} \; (1_{2^{\otimes n-1}} \otimes J_{k} \otimes 1_{2}),
\end{equation}
where only the anyons carrying the defects are shown (the drawing is implicitly preceded by $n-2$ identity strands), the dashed vertical line carries the identity object, and all full black lines carry the object $2$. The spread of vertical black lines on the left and right of the middle dashed line is irrelevant: all choices produces the same generator. The crossings are a shorthand notation for the linear combination\footnote{As our notation suggests, these diagrams satisfy the braid relations and are inverses of each other \cite{braidpaper}.}:
\begin{equation}\label{eq:def.black}
	\begin{tikzpicture}[baseline = {(current bounding box.center)}, every node/.style={scale = 2/3}, scale = 1/2]
	\draw[black, line width = 1pt] (3,1) -- (1,3);
	\draw[white, line width = 3pt] (1,1) -- (3,3);
	\draw[black, line width = 1pt] (1,1) -- (3,3);
	\end{tikzpicture} \; \equiv (-\q)^{1/2}\;
	\begin{tikzpicture}[baseline = {(current bounding box.center)}, every node/.style={scale = 2/3}, scale = 1/2,rotate = 90]
	\draw[black, line width = 1pt] (1,1) -- (2,2) -- (3,1);
	\draw[black, line width = 1pt] (3,4) -- (2,3) -- (1,4);
	\draw[black, dashed, line width = 1pt] (2,3) -- (2,2);
	\end{tikzpicture} \; + (-\q)^{-1/2} \;
	\begin{tikzpicture}[baseline = {(current bounding box.center)}, every node/.style={scale = 2/3}, scale = 1/2]
	\draw[black, line width = 1pt] (1,1) -- (2,2) -- (3,1);
	\draw[black, line width = 1pt] (3,4) -- (2,3) -- (1,4);
	\draw[black, dashed, line width = 1pt] (2,3) -- (2,2);
	\end{tikzpicture}\;,
\end{equation}
\begin{equation}\label{eq:def.white}
	\begin{tikzpicture}[baseline = {(current bounding box.center)}, every node/.style={scale = 2/3}, scale = 1/2]
	\draw[black, line width = 1pt] (1,1) -- (3,3);
	\draw[white, line width = 3pt] (3,1) -- (1,3);
	\draw[black, line width = 1pt] (3,1) -- (1,3);
	\end{tikzpicture} \; \equiv (-\q)^{-1/2}\;
	\begin{tikzpicture}[baseline = {(current bounding box.center)}, every node/.style={scale = 2/3}, scale = 1/2,rotate = 90]
	\draw[black, line width = 1pt] (1,1) -- (2,2) -- (3,1);
	\draw[black, line width = 1pt] (3,4) -- (2,3) -- (1,4);
	\draw[black, dashed, line width = 1pt] (2,3) -- (2,2);
	\end{tikzpicture} \; + (-\q)^{1/2} \;
	\begin{tikzpicture}[baseline = {(current bounding box.center)}, every node/.style={scale = 2/3}, scale = 1/2]
	\draw[black, line width = 1pt] (1,1) -- (2,2) -- (3,1);
	\draw[black, line width = 1pt] (3,4) -- (2,3) -- (1,4);
	\draw[black, dashed, line width = 1pt] (2,3) -- (2,2);
	\end{tikzpicture}\;.
\end{equation}
This deformed generator could thus instead be written as a sum of $2^{2k}$ diagrams.

Similarly, the \emph{under} defect generator can be written
\begin{equation}
\tilde{E}^{2,(k)}_{n-1}[n+k]  \equiv \;
	\begin{tikzpicture}[baseline = {(current bounding box.center)},xscale = 1/2,yscale = 1/3]
	\draw[black, line width = 1pt] (1,9) -- (4,6);
	\draw[black, line width = 1pt] (4,6) -- (7,9);
	\draw[black, line width = 1pt] (1,1) -- (4,4);
	\draw[black, line width = 1pt] (4,4) -- (7,1);
	\draw[black, dashed, line width = 1pt] (4,4) -- (4,6);
	\foreach \s in {2,3,5,6}{
	\draw[white, line width = 3pt] (\s,1) -- (\s,9);
	\draw[black, line width = 1pt] (\s,1) -- (\s,9);
	}
	\foreach \r in {1,...,3}{
		\fill[black] (2 + \r/4 ,3/2 ) circle (1/16);
		\fill[black] (2 + \r/4 ,17/2 ) circle (1/16);
		\fill[black] (5 + \r/4 ,3/2 ) circle (1/16);
		\fill[black] (5 + \r/4 ,17/2 ) circle (1/16);
	}
	\draw[decorate, decoration = {brace, mirror, amplitude = 3 pt}, yshift = -3pt] (1.5,.5) -- (6.5,.5) node [midway,yshift = -7pt] {\footnotesize{k}};
	\end{tikzpicture}  \; (1_{2^{\otimes n-1}} \otimes J_{k} \otimes 1_{2}),
\end{equation}
where only the anyons carrying the defects are shown, and all full black lines carry the object $2$.

\subsubsection{Defects of type $r$}
	Defects of type $r$ are slightly more complicated to describe; the deformed generator is then\footnote{The curious term $\Delta$ is required to preserve the integrability of the model.}:
	\begin{align}
		\mu (\tilde{E}^{2,(k)}_{n} + \Delta) \equiv & \;  w^{-2}\;
		\begin{tikzpicture}[baseline = {(current bounding box.center)},xscale = 1/2,yscale = 1/3]
	\draw[black, line width = 1pt] (1,9) -- (4,6);
	\draw[black, line width = 1pt] (4,6) -- (7,9);
	\draw[black, line width = 1pt] (1,1) -- (4,4);
	\draw[black, line width = 1pt] (4,4) -- (7,1);
	\draw[black, dashed, line width = 1pt] (4,4) -- (4,6);
	\foreach \s in {2,3,5,6}{
	\draw[white, line width = 3pt] (\s,1) -- (\s,9);
	\draw[black, line width = 1pt] (\s,1) -- (\s,9);
	}
	\foreach \r in {1,...,3}{
		\fill[black] (2 + \r/4 ,3/2 ) circle (1/16);
		\fill[black] (2 + \r/4 ,17/2 ) circle (1/16);
		\fill[black] (5 + \r/4 ,3/2 ) circle (1/16);
		\fill[black] (5 + \r/4 ,17/2 ) circle (1/16);
	}
	\draw[decorate, decoration = {brace, mirror, amplitude = 3 pt}, yshift = -3pt] (1.5,.5) -- (6.5,.5) node [midway,yshift = -7pt] {\footnotesize{k}};
	\end{tikzpicture}\; + w^{2} \;
	\begin{tikzpicture}[baseline = {(current bounding box.center)},xscale = 1/2,yscale = 1/3]
	\foreach \s in {2,3,5,6}{
	\draw[black, line width = 1pt] (\s,1) -- (\s,9);
	}
	\draw[white, line width = 3pt] (1,1) -- (4,4);
	\draw[white, line width = 3pt] (4,4) -- (7,1);
	\draw[white, line width = 3pt] (1,9) -- (4,6);
	\draw[white, line width = 3pt] (4,6) -- (7,9);
	\draw[black, line width = 1pt] (1,9) -- (4,6);
	\draw[black, line width = 1pt] (4,6) -- (7,9);
	\draw[black, line width = 1pt] (1,1) -- (4,4);
	\draw[black, line width = 1pt] (4,4) -- (7,1);
	\draw[black, dashed, line width = 1pt] (4,4) -- (4,6);
	\foreach \r in {1,...,3}{
		\fill[black] (2 + \r/4 ,3/2 ) circle (1/16);
		\fill[black] (2 + \r/4 ,17/2 ) circle (1/16);
		\fill[black] (5 + \r/4 ,3/2 ) circle (1/16);
		\fill[black] (5 + \r/4 ,17/2 ) circle (1/16);
	}
	\draw[decorate, decoration = {brace, mirror, amplitude = 3 pt}, yshift = -3pt] (1.5,.5) -- (6.5,.5) node [midway,yshift = -7pt] {\footnotesize{k}};
	\end{tikzpicture}\; \notag\\
	& + \q \;
	\begin{tikzpicture}[baseline = {(current bounding box.center)},xscale = 1/2,yscale = 1/3]
	\foreach \s in {2,3,5,6}{
	\draw[black, line width = 1pt] (\s,4.5) -- (\s,9);
	}
	\draw[white, line width = 3pt] (1,9) -- (4,6);
	\draw[white, line width = 3pt] (4,6) -- (7,9);
	\draw[black, line width = 1pt] (1,9) -- (4,6);
	\draw[black, line width = 1pt] (4,6) -- (7,9);
	\draw[black, line width = 1pt] (1,1) -- (4,4);
	\draw[black, line width = 1pt] (4,4) -- (7,1);
	\draw[black, dashed, line width = 1pt] (4,4) -- (4,6);
	\foreach \s in {2,3,5,6}{
	\draw[white, line width = 3pt] (\s,4.5) -- (\s,1);
	\draw[black, line width = 1pt] (\s,4.5) -- (\s,1);
	}
	\foreach \r in {1,...,3}{
		\fill[black] (2 + \r/4 ,3/2 ) circle (1/16);
		\fill[black] (2 + \r/4 ,17/2 ) circle (1/16);
		\fill[black] (5 + \r/4 ,3/2 ) circle (1/16);
		\fill[black] (5 + \r/4 ,17/2 ) circle (1/16);
	}
	\draw[decorate, decoration = {brace, mirror, amplitude = 3 pt}, yshift = -3pt] (1.5,.5) -- (6.5,.5) node [midway,yshift = -7pt] {\footnotesize{k}};
	\end{tikzpicture} \; + \q^{-1} \;
	\begin{tikzpicture}[baseline = {(current bounding box.center)},xscale = 1/2,yscale = 1/3]
	\draw[black, line width = 1pt] (1,9) -- (4,6);
	\draw[black, line width = 1pt] (4,6) -- (7,9);	
	\foreach \s in {2,3,5,6}{
	\draw[white, line width = 3pt] (\s,1) -- (\s,9);
	\draw[black, line width = 1pt] (\s,1) -- (\s,9);
	}
	\draw[white, line width = 3pt] (1,1) -- (4,4);
	\draw[white, line width = 3pt] (4,4) -- (7,1);
	\draw[black, line width = 1pt] (1,1) -- (4,4);
	\draw[black, line width = 1pt] (4,4) -- (7,1);
	\draw[black, dashed, line width = 1pt] (4,4) -- (4,6);
	\foreach \r in {1,...,3}{
		\fill[black] (2 + \r/4 ,3/2 ) circle (1/16);
		\fill[black] (2 + \r/4 ,17/2 ) circle (1/16);
		\fill[black] (5 + \r/4 ,3/2 ) circle (1/16);
		\fill[black] (5 + \r/4 ,17/2 ) circle (1/16);
	}
	\draw[decorate, decoration = {brace, mirror, amplitude = 3 pt}, yshift = -3pt] (1.5,.5) -- (6.5,.5) node [midway,yshift = -7pt] {\footnotesize{k}};
	\end{tikzpicture} \;,
	\end{align}
	\begin{equation}
	\Delta = (\q^{2}+\q^{-2})\frac{w-w^{-1}}{w+w^{-1}}1_{2^{\otimes n +k}} - \q \;
	\begin{tikzpicture}[baseline = {(current bounding box.center)},xscale = 1/2,yscale = 1/3]
	\draw[black, line width = 1pt] (0,3) .. controls  (0,4) and (4,4) .. (4,5);
	\foreach \r in {1,3}{
		\draw[white, line width = 3pt] (\r,1) -- (\r,5);
		\draw[black, line width = 1pt] (\r,1) -- (\r,5);
	};
	\draw[white, line width = 3pt] (4,1) .. controls  (4,2) and (0,2) .. (0,3);
	\draw[black, line width = 1pt] (4,1) .. controls  (4,2) and (0,2) .. (0,3);
	\foreach \r in {-1,0,1}{
		\fill[black] (2 + \r/2 , 5 ) circle (1/16);
		\fill[black] (2 + \r/2 , 1 ) circle (1/16);
	}
	\draw[decorate, decoration = {brace, mirror, amplitude =2 pt}, yshift = -2pt] (0.5,.5) -- (3.5,.5) node [midway,yshift = -7pt] {\footnotesize{k}};
	\end{tikzpicture} \; + \q^{-1} \;
	\begin{tikzpicture}[baseline = {(current bounding box.center)},xscale = 1/2,yscale = 1/3]
	\draw[black, line width = 1pt] (4,1) .. controls  (4,2) and (0,2) .. (0,3);
	\foreach \r in {1,3}{
		\draw[white, line width = 3pt] (\r,1) -- (\r,5);
		\draw[black, line width = 1pt] (\r,1) -- (\r,5);
	};
	\draw[white, line width = 3pt] (0,3) .. controls  (0,4) and (4,4) .. (4,5);
	\draw[black, line width = 1pt] (0,3) .. controls  (0,4) and (4,4) .. (4,5);
	\foreach \r in {-1,0,1}{
		\fill[black] (2 + \r/2 , 5 ) circle (1/16);
		\fill[black] (2 + \r/2 , 1 ) circle (1/16);
	}
	\draw[decorate, decoration = {brace, mirror, amplitude = 2 pt}, yshift = -2pt] (0.5,.5) -- (3.5,.5) node [midway,yshift = -7pt] {\footnotesize{k}};
	\end{tikzpicture}
	\end{equation}
	\begin{equation*}
	\mu = (-\q)^{k+1} + (-\q)^{-k-1}-w^{2} - w^{-2}
	\end{equation*}
which is further multiplied by the Jones-Wenzl projector, just like the two defects of type $s$. The parameter $w$ is the spectral parameter of the defect, which can be any non-zero real number\footnote{If one allows $w$ to have a non-trivial imaginary part, this classification is redundant: defects of large width can be obtained by putting a complex $w$ in a smaller defect. See \cite{nextpaper}.}; in the limit where it goes to zero, or infinity, one recovers the expression for the defect of type $s$. See section \ref{sec:topdefect.analytics} for more details.
In terms of Temperley-Lieb generators, the deformed generator corresponding to a defect of width $k$ sitting at position $j+1$ can be expressed as
\begin{align}
	\tilde{e}_{j}^{(k)} = & \frac{\q-\q^{-1}}{2}\lambda + e_{j} +e_{j+k} - (1-\delta_{k,1})(\q +\q^{-1})\alpha_{k-1,0}e_{j}e_{j+k} \notag \\
	& + \alpha_{1,-k}e_{j}e_{j+1}\hdots e_{j+k} +\alpha_{1,k}e_{j+k}e_{j+k-1}\hdots e_{j}  \;,
\end{align}
which must further be multiplied on both sides by the appropriate idempotent, and where
\begin{align*}
	\alpha_{x,y} &= \frac{(-\q)^{x} + (-\q)^{-x} - (-\q)^{y}w^{2} - (-\q)^{-y}w^{-2}}{(-\q)^{k+1} + (-\q)^{-k-1} - w^{2} - w^{-2}} \;,\\
	\lambda & = 2 \frac{1+(-\q)^{1+k}}{(w+w^{-1})(w^{-1}- (-\q)^{k+1}w)}.
\end{align*}

\subsection{Defects in the crossed channel}

In the crossed channel, defects appear as a family of elements of the Temperley-Lieb algebras acting on the Hilbert space by simple multiplication. As in the direct channel, one can break these into three distinct family of elements; we start with the $s$-type and then move on to the $r$-type.
\subsubsection{Defects of type $s$}
The $s$-type defects appear as central elements of $\atl{n}$ which we usually refer to as the \emph{hoop} operators (see \cite{BGJST}). In the anyonic formulation, they can be written \footnote{These defects will turn out to be equivalent to those built in \cite{BuicanGromov}.}
\begin{equation}
	Y_{a} \; \equiv \;
	\begin{tikzpicture}[baseline = {(current bounding box.center)}, scale = 1/2]
	\foreach \r in {1,2,4,5}{
		\draw[black, line width = 1pt] (\r,1) -- (\r,3);
		\node[anchor = south,style={scale = 2/3}] at (\r,3) {$a$};
		\node[anchor = north,style={scale = 2/3}] at (\r,1) {$a$};
	}
	\draw[white, line width = 3pt] (0,2) -- (6,2);
	\draw[black, line width = 1pt] (0,2) -- (6,2);
	\node[anchor = south,style={scale = 2/3}] at (0.5,2) {$a$};
	\node[anchor = south] at (3,2) {$\hdots $};
	\node[anchor = north] at (3,2) {$\hdots $};
	\draw[black, dashed, line width = 1pt] (0,1) -- (0,3);
	\draw[black, dashed, line width = 1pt] (6,1) -- (6,3);
	\end{tikzpicture} \;,
\end{equation}
\begin{equation}
	\bar{Y}_{a} \; \equiv \;
	\begin{tikzpicture}[baseline = {(current bounding box.center)}, scale = 1/2]
	\draw[black, line width = 1pt] (0,2) -- (6,2);
	\foreach \r in {1,2,4,5}{
		\draw[white, line width = 3pt] (\r,1) -- (\r,3);
		\draw[black, line width = 1pt] (\r,1) -- (\r,3);
		\node[anchor = south,style={scale = 2/3}] at (\r,3) {$a$};
		\node[anchor = north,style={scale = 2/3}] at (\r,1) {$a$};
	}
	\node[anchor = south,style={scale = 2/3}] at (0.5,2) {$a$};
	\node[anchor = south] at (3,2) {$\hdots $};
	\node[anchor = north] at (3,2) {$\hdots $};
	\draw[black, dashed, line width = 1pt] (0,1) -- (0,3);
	\draw[black, dashed, line width = 1pt] (6,1) -- (6,3);
	\end{tikzpicture} \;,
\end{equation}
where the two opposing dashed lines are identified, and the braidings were introduced in equations \eqref{eq:def.black} and \eqref{eq:def.white}. One can show that these are indeed (the image of) central elements of the affine Temperley-Lieb algebra, so they commute with each individual $E^{a}_{i}[n]$, and therefore with the Hamiltonian \eqref{eq:anyonHamil} itself. Richer defect operators can then be built by considering polynomials in these two operators (see \cite{BGJST}).

For the RSOS models, where $a=2$, the states in equation \eqref{eq:basis.hilbert} were specifically chosen so that the action of these defect operators is easy to study:
\begin{align}\label{eq:hoopcalculation}
	Y_{2}U_{r}[n] \;& = \;
	\begin{tikzpicture}[baseline = {(current bounding box.center)}, every node/.style={scale = 1}, scale = 2/3]
	\foreach \r in {2,3,4,5,8,9}{
	\node (a\r) at (\r,3) {$\footnotesize{2}$};
	}
	\foreach \r in {5,9,17}{
	\draw[black, dashed, line width = 1pt] (\r/2,2) -- (\r/2,1);
	}
	\draw[black, line width = 1pt] (a2) -- (5/2,2);
	\draw[black, line width = 1pt] (a3) -- (5/2,2);
	\draw[black, line width = 1pt] (a4) -- (9/2,2);
	\draw[black, line width = 1pt] (a5) -- (9/2,2);
	\draw[black, line width = 1pt] (a8) -- (17/2,2);
	\draw[black, line width = 1pt] (a9) -- (17/2,2);
	\node (a0) at (1,1) {$\footnotesize{r}$};
	\node (a10) at (10,1) {$\footnotesize{r}$};
	\node[anchor = north] at (7/2,1) {$\footnotesize{r}$};
	\node[anchor = north] at (11/2,1) {$\footnotesize{r}$};
	\node[anchor = north] at (15/2,1) {$\footnotesize{r}$};
	\node[anchor = north] at (13/2,2) {$\hdots$};
	\draw[black, line width = 1pt] (a0) -- (a10);
	\foreach \r in {2,3,4,5,8,9}{
		\draw[black, line width = 1pt] (a\r) -- (\r,5);
		\node[anchor = south] at (\r,5) {$2$};
	}
	\draw[white, line width = 3pt] (1,4) -- (10,4);
	\draw[black, line width = 1pt] (1,4) -- (10,4);
	\node[anchor = south] at (1.5,4) {$2$};
	\node[anchor = south] at (6.5,4) {$\hdots $};
	\node[anchor = north] at (6.5,4) {$\hdots $};
	\end{tikzpicture} \; \notag\\
	& = \;
	\begin{tikzpicture}[baseline = {(current bounding box.center)}, every node/.style={scale = 1}, scale = 2/3]
	\foreach \r in {2,3,4,5,8,9}{
	\node (a\r) at (\r,3) {$\footnotesize{2}$};
	}
	\foreach \r in {5,9,17}{
	\draw[black, dashed, line width = 1pt] (\r/2,2) -- (\r/2,1);
	}
	\draw[black, line width = 1pt] (a2) -- (5/2,2);
	\draw[black, line width = 1pt] (a3) -- (5/2,2);
	\draw[black, line width = 1pt] (a4) -- (9/2,2);
	\draw[black, line width = 1pt] (a5) -- (9/2,2);
	\draw[black, line width = 1pt] (a8) -- (17/2,2);
	\draw[black, line width = 1pt] (a9) -- (17/2,2);
	\node (a0) at (1,1) {$\footnotesize{r}$};
	\node (a10) at (10,1) {$\footnotesize{r}$};
	\node[anchor = north] at (7/2,1) {$\footnotesize{r}$};
	\node[anchor = north] at (11/2,1) {$\footnotesize{r}$};
	\node[anchor = north] at (15/2,1) {$\footnotesize{r}$};
	\node[anchor = north] at (13/2,2) {$\hdots$};
	\draw[black, line width = 1pt] (a0) -- (a10);
	\draw[black, line width = 1pt] (1,0) -- (10,0);
	\draw[black, dashed, line width = 1pt] (3,0) -- (3,1);
	\node[anchor = east] at (1,0) {$2$};
	\end{tikzpicture}  \; \notag \\
	& =  \sum_{y} \frac{(\mathsf{F}_{2,r,r}^{2})^{1}_{y}}{(\mathsf{F}_{r,r,2}^{2})^{y}_{1}} \left((\mathsf{F}_{2,r,1}^{y})^{r}_{y} \right)^{n/2} U_{r}[n] \notag \\
	& = U_{r-1}[n] + U_{r+1}[n],
\end{align}
where we used our conventions for the $F$-symbols (see appendix \ref{sec:fsymbols}) to simplify the last line, and we assume that $U_{0}[n] \equiv U_{p+1}[n] \equiv 0 $. It follows in particular that the eigenvalues of $Y_{2}$ are
	\begin{equation}\label{eq:hoop.eigenvalues}
		\left\lbrace 2 \cos(\frac{\pi}{p+1}),2 \cos(\frac{2\pi}{p+1}), \hdots, 2 \cos(\frac{p\pi}{p+1}) \right\rbrace \;.
	\end{equation}
It is easy to check  that this formulation is identical to the one in \cite{Gilsetal,Feiguinetal}\footnote{$Y_2$ here is denoted by $Y_\tau$ there. More generally, our conventions are such that $a$ corresponds to an $SU(2)$ spin-$l$ with $a=2l+1$.}. In other words, the ``topological symmetry'' operators are those generating the center of the affine Temperley-Lieb algebra.

A similar calculation shows that
\begin{equation}
	\bar{Y}_{2}U_{r}[n] = Y_{2}U_{r}[n].
\end{equation}
This coincidence of the two defect operators happens because we chose periodic boundary conditions; when one considers twisted boundary conditions, so the fusion trees do not have to wrap neatly on themselves, these two defects operators must be computed separately, and usually do not coincide.
\subsubsection{Defects of type $r$}
The last family of defect operators can be defined as
\begin{equation}
	\mathbb{Y}_{k}(x) \equiv \;
	\begin{tikzpicture}[baseline = {(current bounding box.center)}, every node/.style={scale = 2/3}, scale = 1/2]
	\draw[black, line width = 1pt] (1,2) -- (10,2);
	\draw[black, line width = 1pt] (1,4) -- (10,4);
	\foreach \r in {2,4,6,8}{
		\draw[black, line width = 1pt] (\r,1) -- (\r,5);
		\fill[white] (\r,2) circle (1/8);
		\fill[white] (\r,4) circle (1/8);
		\draw[black,line width = 1pt] (\r,2) circle (1/8);
		\draw[black,line width = 1pt] (\r,4) circle (1/8);
		\fill[gray] (\r,2) circle (3/32);
		\fill[gray] (\r,4) circle (3/32);
		\node[anchor = south] at (\r,5) {$a$};
		\node[anchor = north] at (\r,1) {$a$};
	}
	\fill[white] (2.5,1.5) -- (3.5,1.5) -- (3.5,4.5) -- (2.5,4.5) -- (2.5,1.5);
	\draw[black, line width = 1pt] (2.5,1.5) -- (3.5,1.5) -- (3.5,4.5) -- (2.5,4.5) -- (2.5,1.5);
	\node at (3,3) {$k$};
	\fill[white] (4.5,1.5) -- (5.5,1.5) -- (5.5,4.5) -- (4.5,4.5) -- (4.5,1.5);
	\draw[black, line width = 1pt] (4.5,1.5) -- (5.5,1.5) -- (5.5,4.5) -- (4.5,4.5) -- (4.5,1.5);
	\node at (5,3) {$k$};
	\fill[white] (8.5,1.5) -- (9.5,1.5) -- (9.5,4.5) -- (8.5,4.5) -- (8.5,1.5);
	\draw[black, line width = 1pt] (8.5,1.5) -- (9.5,1.5) -- (9.5,4.5) -- (8.5,4.5) -- (8.5,1.5);
	\node at (9,3) {$k$};
	\fill[white] (6.5,1.5) -- (7.5,1.5) -- (7.5,4.5) -- (6.5,4.5) -- (6.5,1.5);
	\foreach \r in {1,2,3}{
	\fill[black] (1.5,2+\r/2) circle (1/16);
	\fill[black] (9.75,2+\r/2) circle (1/16);
	\fill[black] (6+ \r/2,3) circle (1/16);
	}
	\node[anchor = north] at (1.25,2) {$a$};
	\node[anchor = south] at (1.25,4) {$a$};
	\draw[black, dashed, line width = 1pt] (1,1) -- (1,5);
	\draw[black, dashed, line width = 1pt] (10,1) -- (10,5);
	\end{tikzpicture}\;,
\end{equation}
where there are $k$ horizontal lines, the white boxes are primitive idempotents acting horizontally, and the gray vertices are a shorthand notation for the linear combination
	\begin{equation}\label{eq:def.graydot}
	\begin{tikzpicture}[baseline = {(current bounding box.center)}, every node/.style={scale = 2/3}, scale = 1/2,rotate = - 45]
	\draw[black, line width = 1pt] (1,1) -- (3,3);
	\draw[black, line width = 1pt] (3,1) -- (1,3);
	\draw[black, line width = 1pt] (2,2) circle (1/8);
	\fill[gray] (2,2) circle (3/32);
	\end{tikzpicture} \; \equiv x (-\q)^{-1/2}\;
	\begin{tikzpicture}[baseline = {(current bounding box.center)}, every node/.style={scale = 2/3}, scale = 1/2,rotate = -45]
	\draw[black, line width = 1pt] (3,1) -- (1,3);
	\draw[white, line width = 3pt] (1,1) -- (3,3);
	\draw[black, line width = 1pt] (1,1) -- (3,3);
	\end{tikzpicture}  \; - x^{-1} (-\q)^{1/2} \;
	\begin{tikzpicture}[baseline = {(current bounding box.center)}, every node/.style={scale = 2/3}, scale = 1/2, rotate = -45]
	\draw[black, line width = 1pt] (1,1) -- (3,3);
	\draw[white, line width = 3pt] (3,1) -- (1,3);
	\draw[black, line width = 1pt] (3,1) -- (1,3);
	\end{tikzpicture}\;.
\end{equation}
For the RSOS models, which corresponds to $a =2$, this expression can be simplified further: one can show that
\begin{equation}
	\begin{tikzpicture}[baseline = {(current bounding box.center)}, every node/.style={scale = 2/3}, scale = 1/2]
	\draw[black, line width = 1pt] (2,2) -- (6,2);
	\draw[black, line width = 1pt] (2,4) -- (6,4);
	\foreach \r in {4}{
		\draw[black, line width = 1pt] (\r,1) -- (\r,5);
		\fill[white] (\r,2) circle (1/8);
		\fill[white] (\r,4) circle (1/8);
		\draw[black,line width = 1pt] (\r,2) circle (1/8);
		\draw[black,line width = 1pt] (\r,4) circle (1/8);
		\fill[gray] (\r,2) circle (3/32);
		\fill[gray] (\r,4) circle (3/32);
		\node[anchor = south] at (\r,5) {$2$};
		\node[anchor = north] at (\r,1) {$2$};
	}
	\fill[white] (2.5,1.5) -- (3.5,1.5) -- (3.5,4.5) -- (2.5,4.5) -- (2.5,1.5);
	\draw[black, line width = 1pt] (2.5,1.5) -- (3.5,1.5) -- (3.5,4.5) -- (2.5,4.5) -- (2.5,1.5);
	\node at (3,3) {$k$};
	\fill[white] (4.5,1.5) -- (5.5,1.5) -- (5.5,4.5) -- (4.5,4.5) -- (4.5,1.5);
	\draw[black, line width = 1pt] (4.5,1.5) -- (5.5,1.5) -- (5.5,4.5) -- (4.5,4.5) -- (4.5,1.5);
	\node at (5,3) {$k$};
	\foreach \r in {1,2,3}{
	\fill[black] (2.25,2+\r/2) circle (1/16);
	\fill[black] (5.75,2+\r/2) circle (1/16);
	}
	\end{tikzpicture}\; = \omega_{1}
	\;\begin{tikzpicture}[baseline = {(current bounding box.center)}, every node/.style={scale = 2/3}, scale = 1/2]
	\foreach \r in {4}{
		\draw[black, line width = 1pt] (\r,1) -- (\r,5);
		\node[anchor = south] at (\r,5) {$2$};
		\node[anchor = north] at (\r,1) {$2$};
	}
	\draw[white, line width = 3pt] (2,2) -- (6,2);
	\draw[white, line width = 3pt] (2,4) -- (6,4);
	\draw[black, line width = 1pt] (2,2) -- (6,2);
	\draw[black, line width = 1pt] (2,4) -- (6,4);
	\fill[white] (2.5,1.5) -- (3.5,1.5) -- (3.5,4.5) -- (2.5,4.5) -- (2.5,1.5);
	\draw[black, line width = 1pt] (2.5,1.5) -- (3.5,1.5) -- (3.5,4.5) -- (2.5,4.5) -- (2.5,1.5);
	\node at (3,3) {$k$};
	\fill[white] (4.5,1.5) -- (5.5,1.5) -- (5.5,4.5) -- (4.5,4.5) -- (4.5,1.5);
	\draw[black, line width = 1pt] (4.5,1.5) -- (5.5,1.5) -- (5.5,4.5) -- (4.5,4.5) -- (4.5,1.5);
	\node at (5,3) {$k$};
	\foreach \r in {1,2,3}{
	\fill[black] (2.25,2+\r/2) circle (1/16);
	\fill[black] (5.75,2+\r/2) circle (1/16);
	}
	\end{tikzpicture}\; + \omega_{2} \;
	\begin{tikzpicture}[baseline = {(current bounding box.center)}, every node/.style={scale = 2/3}, scale = 1/2]
	\draw[black, line width = 1pt] (2,2) -- (6,2);
	\draw[black, line width = 1pt] (2,4) -- (6,4);
	\foreach \r in {4}{
		\draw[white, line width = 3pt] (\r,1) -- (\r,5);
		\draw[black, line width = 1pt] (\r,1) -- (\r,5);
		\node[anchor = south] at (\r,5) {$2$};
		\node[anchor = north] at (\r,1) {$2$};
	}
	\fill[white] (2.5,1.5) -- (3.5,1.5) -- (3.5,4.5) -- (2.5,4.5) -- (2.5,1.5);
	\draw[black, line width = 1pt] (2.5,1.5) -- (3.5,1.5) -- (3.5,4.5) -- (2.5,4.5) -- (2.5,1.5);
	\node at (3,3) {$k$};
	\fill[white] (4.5,1.5) -- (5.5,1.5) -- (5.5,4.5) -- (4.5,4.5) -- (4.5,1.5);
	\draw[black, line width = 1pt] (4.5,1.5) -- (5.5,1.5) -- (5.5,4.5) -- (4.5,4.5) -- (4.5,1.5);
	\node at (5,3) {$k$};
	\foreach \r in {1,2,3}{
	\fill[black] (2.25,2+\r/2) circle (1/16);
	\fill[black] (5.75,2+\r/2) circle (1/16);
	}
	\end{tikzpicture}\; ,
\end{equation}
where $\omega_{1}, \omega_{2}$ are functions of $x$ which also depends on $k$ and the specific choice of idempotent for the white boxes. Note in particular that equivalent idempotents ($\mu, \nu$ such that $\tl{n}\mu \simeq \tl{n}\nu $) do not necessarily yield the same functions $\omega_{1}, \omega_{2}$.

\subsection{What is \emph{topological} about these defects?}

\begin{figure}
\begin{center}
\begin{tikzpicture}[baseline = {(current bounding box.center)}, every node/.style={scale = 1/2}, scale = 2/3]
	\draw[black, line width = 1pt] (1,1) -- (1,3);
	\draw[black, line width = 1pt] (2,1) -- (2.5,1.5) -- (3,1);
	\draw[black, line width = 1pt] (2,3) -- (2.5,2.5) -- (3,3);
	\draw[black, line width = 1pt] (4,1) -- (4,3);
	\draw[black, dashed, line width = 1pt] (2.5,1.5) -- (2.5,2.5);
	\end{tikzpicture} \; $\sim$  \;
\begin{tikzpicture}[baseline = {(current bounding box.center)}, every node/.style={scale = 1/2}, scale = 2/3]
	\draw[black, line width = 1pt] (1,1) -- (1,3);
	\draw[black, line width = 1pt] (2,1) -- (2.5,1.5) -- (4,1);
	\draw[black, line width = 1pt] (2,3) -- (2.5,2.5) -- (4,3);
	\draw[white, line width = 3pt] (3,1) -- (3,3);
	\draw[black, line width = 1pt] (3,1) -- (3,3);
	\draw[black, dashed, line width = 1pt] (2.5,1.5) -- (2.5,2.5);
	\end{tikzpicture} \; $\sim$  \;
\begin{tikzpicture}[baseline = {(current bounding box.center)}, every node/.style={scale = 1/2}, scale = 2/3]
	\draw[black, line width = 1pt] (1,1) -- (1,3);
	\draw[black, line width = 1pt] (2,1) -- (3.5,1.5) -- (4,1);
	\draw[black, line width = 1pt] (2,3) -- (3.5,2.5) -- (4,3);
	\draw[white, line width = 3pt] (2.5,1) -- (2.5,3);
	\draw[black, line width = 1pt] (2.5,1) -- (2.5,3);
	\draw[black, dashed, line width = 1pt] (3.5,1.5) -- (3.5,2.5);
	\end{tikzpicture} \; $\sim$ \;
\begin{tikzpicture}[baseline = {(current bounding box.center)}, every node/.style={scale = 1/2}, scale = 2/3]
	\draw[black, line width = 1pt] (1,1) -- (1,3);
	\draw[black, line width = 1pt] (3,1) -- (3.5,1.5) -- (4,1);
	\draw[black, line width = 1pt] (3,3) -- (3.5,2.5) -- (4,3);
	\draw[white, line width = 3pt] (2,1) -- (2,3);
	\draw[black, line width = 1pt] (2,1) -- (2,3);
	\draw[black, dashed, line width = 1pt] (3.5,1.5) -- (3.5,2.5);
	\end{tikzpicture}
\end{center}
\caption{Moving a defect in the direct channel through anyon fusions; unmarked black lines carry the object $2$ while the dashed lines carry the identity object. The resulting operators are all equivalent up to a local change of basis, that is operators who do not act on the four depicted lines will be left unchanged by the change of basis.}\label{fig:direct}
\end{figure}
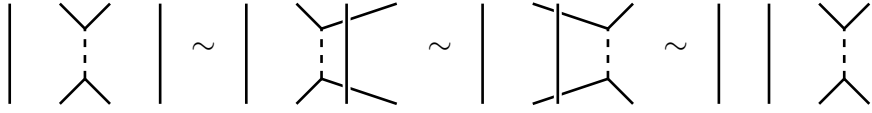

\begin{figure}
	\begin{center}
	\begin{tikzpicture}[baseline = {(current bounding box.center)}, every node/.style={scale = 1/2}, scale = 2/3]
	\draw[black, line width = 1pt] (1,1) -- (1,3);
	\draw[black, line width = 1pt] (2,1) -- (2.5,1.5) -- (3,1);
	\draw[black, line width = 1pt] (2,3) -- (2.5,2.5) -- (3,3);
	\draw[black, line width = 1pt] (4,1) -- (4,3);
	\draw[black, dashed, line width = 1pt] (2.5,1.5) -- (2.5,2.5);
	\foreach \r in {1,2,3,4}{
		\draw[black, line width = 1pt] (\r,3) -- (\r,4);
	};
	\draw[white, line width = 3pt] (.5,3.5) -- (4.5,3.5);
	\draw[black, line width = 1pt] (.5,3.5) -- (4.5,3.5);
	\end{tikzpicture} \; = \;
	\begin{tikzpicture}[baseline = {(current bounding box.center)}, every node/.style={scale = 1/2}, scale = 2/3]
	\draw[black, line width = 1pt] (1,1) -- (1,3);
	\draw[black, line width = 1pt] (2,1) -- (2.5,1.5) -- (3,1);
	\draw[black, line width = 1pt] (2,3) -- (2.5,2.5) -- (3,3);
	\draw[black, line width = 1pt] (4,1) -- (4,3);
	\draw[black, dashed, line width = 1pt] (2.5,1.5) -- (2.5,2.5);
	\foreach \r in {1,2,3,4}{
		\draw[black, line width = 1pt] (\r,0) -- (\r,1);
	};
	\draw[white, line width = 3pt] (.5,0.5) -- (4.5,0.5);
	\draw[black, line width = 1pt] (.5,0.5) -- (4.5,0.5);
	\end{tikzpicture} \; \qquad \;
	\begin{tikzpicture}[baseline = {(current bounding box.center)}, every node/.style={scale = 1/2}, scale = 2/3]
	\draw[black, line width = 1pt] (1,1) -- (1.5,.5) -- (2,1);
	\draw[black, line width = 1pt] (1.5,.5) -- (2,0) -- (3,1);
	\draw[black, line width = 1pt] (2,0) -- (2,-1);
	\foreach \r in {1,2,3}{
		\draw[black, line width = 1pt] (\r,1) -- (\r,2);
	};
	\draw[white, line width = 3pt] (.5,1.5) -- (3.5,1.5);
	\draw[black, line width = 1pt] (.5,1.5) -- (3.5,1.5);
	\node[anchor = south] at (1,2) {$a$};
	\node[anchor = south] at (2,2) {$b$};
	\node[anchor = south] at (3,2) {$c$};
	\node[anchor = north] at (2,-1) {$d$};
	\node[anchor = east] at (1.75,0) {$f$};
	\end{tikzpicture} \; = \;
	\begin{tikzpicture}[baseline = {(current bounding box.center)}, every node/.style={scale = 1/2}, scale = 2/3]
	\draw[black, line width = 1pt] (1,1) -- (1.5,.5) -- (2,1);
	\draw[black, line width = 1pt] (1.5,.5) -- (2,0) -- (3,1);
	\draw[black, line width = 1pt] (2,0) -- (2,-1);
	\foreach \r in {2}{
		\draw[black, line width = 1pt] (\r,-1) -- (\r,-2);
	};
	\draw[white, line width = 3pt] (.5,-1.5) -- (3.5,-1.5);
	\draw[black, line width = 1pt] (.5,-1.5) -- (3.5,-1.5);
	\node[anchor = south] at (1,1) {$a$};
	\node[anchor = south] at (2,1) {$b$};
	\node[anchor = south] at (3,1) {$c$};
	\node[anchor = north] at (2,-2) {$d$};
	\node[anchor = east] at (1.75,0) {$f$};
	\end{tikzpicture}
	\end{center}
\caption{Moving a defect in the cross channel through anyon fusions; unmarked black lines carry the object $2$ while the dashed lines carry the identity object. The defect line commutes with all fusions, even those not preserving the total number of anyons.}\label{fig:cross}
\end{figure}
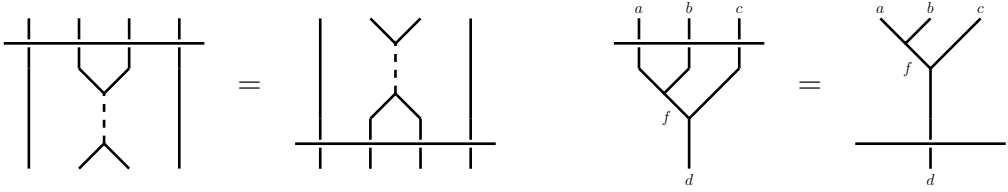

The defects of type $s$ are, in a natural sense, topological ``on the lattice'', that is, the lines carrying them on the cylinder can be deformed without affecting any of their physical properties \cite{BGJST}. This Euclidian version of what it means to be topological translates into different formulations in the direct (figure \ref{fig:direct}) and crossed channel (figure \ref{fig:cross}). In the crossed channel, in particular, we find below that the defect operators commute with the lattice regularizations of the Virasoro generators \cite{KooSaleur,Vidal}, ensuring that their continuum limit satisfies (\ref{VirComm}) indeed.

\subsubsection{$s$-type in the direct channel}
The \emph{clean} Hamiltonian (without any defect)
\begin{equation*}
	H_{2}[n] = \sum_{i = 1}^{n} E^{2}_{i}[n],
\end{equation*}
is manifestly homogeneous since the coefficient in front of each Temperley-Lieb generator $E^{2}_{i}[n]$ is the same; replacing any one of these with one of our deformed generators will thus obviously produce equivalent Hamiltonians: this is not topological invariance, merely translation invariance. However, the $s$-type defects enjoy a much larger symmetry \cite{BGJST}:

\emph{
Let $\mathfrak{H}_{2}[n] \in \mathsf{End}_{\mathcal{C}_{p}}(2^{\otimes n})$ be some morphism written as a finite sum of finite products of the generators $E^{2}_{i}[n]$, $ i =1, \hdots, n$. Let $\mathfrak{H}^{(k)}_{2}[n;i]$ be the endomorphism of $2^{\otimes (n + k)}$ obtained by replacing all instances of the generator $E^{2}_{i}[n]$ with a deformed generator of $s$-type, of width $k$, and sending
\begin{equation*}
	E^{2}_{j}[n] \to (1_{2^{\otimes i}} \otimes J_{k} \otimes 1_{2^{\otimes n-i}})\left. \begin{cases}
		E^{2}_{j}[n+k] & \text{if } 0 < j < i\\
		E^{2}_{j + k}[n+k] & \text{if } i < j \leq n\\
	\end{cases}\right\rbrace,
\end{equation*}
where $J_{k}$ is an idempotent projecting on the anyon $k$ (compare these expressions with \eqref{def:vanilladefect}). Then
\begin{equation}
	\mathfrak{H}^{(k)}_{2}[n;i] \text{ is similar to } \mathfrak{H}^{(k)}_{2}[n;j] \qquad \forall i,j = 1, \hdots, n,
\end{equation}
as matrices.
}
In other words,  the spectrum of a general, non-homogenous, Temperley-Lieb  Hamiltonian with a $s$-type defect does not depend on where this defect is introduced. Note that this result does not assume that $E^{2}_{i}[n]$ appears at all in $\mathfrak{H}_{2}[n]$!

\subsubsection{$s$-type in the crossed channel}

Since the two defect operators $Y_{2}[n], \bar{Y}_{2}[n]$ are central in the affine Temperley-Lieb algebra, and that the Hilbert space of the periodic RSOS model is a semi-simple $\atl{n}$-module, if there exists some lattice realization of the Virasoro generators $\mathsf{L}^{(n)}_{m}, \bar{\mathsf{L}}^{(n)}_{m} $ as endomorphisms of $2^{\otimes n}$ then
\begin{equation*}
	\big[\mathsf{L}^{(n)}_{m},Y_{2}[n]\big] = \big[\mathsf{L}^{(n)}_{m},\bar{Y}_{2}[n]\big] = 0, \qquad \big[\bar{\mathsf{L}}_{m}^{(n)},Y_{2}[n]\big] = \big[\bar{\mathsf{L}}^{(n)}_{m},\bar{Y}_{2}[n]\big] = 0.
\end{equation*}
Of course, commutation for the lattice model implies commutation in the continuum limit as well, proving that (\ref{VirComm}) holds indeed. However, one can make a slightly stronger statement:

\emph{
Let $n,m >0 $ be integers, and $F_{m,n} \in \mathsf{Hom}_{\mathcal{C}_{p}}(2^{\otimes n}, 2^{\otimes m})$. Then
\begin{equation}
	F_{m,n}Y_{2}[n] = Y_{2}[m]F_{m,n}, \qquad F_{m,n}\bar{Y}_{2}[n] = \bar{Y}_{2}[m]F_{m,n},
\end{equation}
that is, the defect operators are the components of a natural endomorphism of the identity functor on the subcategory of $\mathcal{C}_{p}$ with objects of the form $\lbrace 2^{\otimes n}| n = 0,1,2 \hdots\rbrace$, with it being understood that $2^{\otimes 0} \equiv 1 $.
}
In other words the hoop operators also commute with linear maps which do not preserve the number of anyons in the chain;
in particular this means that the eigenvalues of the defect operators are independent of the size of the chain.

\subsection{$r$-type defects}

The $r$-type defects we have proposed in the previous section are clearly not topological on the lattice: properties depend on where exactly they are placed, and, if one looks for instance at the crossed channel, the corresponding operator \emph{does not commute} with the lattice Virasoro generators.  However, analytical and numerical test shows that they should become topological in the continuum limit: this is discussed in part below, and more fully in \cite{nextpaper}.

\section{Merging of $s$-type defects}\label{sec:topdefect.merging}
While the previous section describes how to construct Hamiltonians carrying a single defect, one might also want to construct ones carrying multiple defects in different channels. However, it turns out that every such models can be reduced, up to a change of basis, to one carrying a single defect; this section explains how this reduction is obtained, focussing on the 
''under" defects. The merging of ''over" defects is obtained in a very similar way, and the two classes of defects do not interact with each others, as long as they are in the same channel.
\subsection{Direct channel}
Recall that a Hamiltonian carrying a single defect of type $s$ in the direct channel can be constructed by simply replacing one of the Temperley-Lieb generators with a \emph{deformed} one, as in equation \eqref{eq:hamilt.1defect.direct}. Multiple defects can be added by simply adding more deformed generators multiplied by the appropriate idempotents; one can show that the position of those new deformed generators (their bottom index) is irrelevant, as long as they do not overlap: the resulting Hamiltonians are all equivalent up to a change of basis. One can also show that instead of adding $m$ different new deformed generators, one could simply add a single \emph{multi-deformed} one, to get the Hamiltonian
\begin{equation}
\tilde{H}^{(k_1,k_2,\hdots, k_{m})}_{2}[n] \sim \sum_{i = 0}^{n-2} \tilde{E}^{2}_{i}[n+K] + \tilde{E}^{2, (k_{1},k_{2},\hdots, k_{m})}_{n-1}[n+K],
\end{equation}
where $\lbrace k_{j} \rbrace_{j=1,\hdots,m}$ is the set of the width of all the defects, $K = \sum_{j}k_{j}$ is their total width; the deformed generators are, for $i\neq n-1 $,
\begin{equation}
	\tilde{E}^{2}_{i}[n+K] = E^{2}_{i}[n+K](1_{2^{\otimes n-1}} \otimes J_{k_{1}} \otimes J_{k_{2}} \otimes \hdots \otimes J_{k_{m}} \otimes 1_{2}),
\end{equation}
and
\begin{equation}
\tilde{E}^{2,\lbrace k_{j}\rbrace}_{n-1}[n+K]  \equiv \;
	\begin{tikzpicture}[baseline = {(current bounding box.center)},xscale = 1/2,yscale = 1/3]
	\foreach \s in {2,3,5,6}{
	\draw[black, line width = 1pt] (\s,1) -- (\s,9);
	}
	\draw[white, line width = 3pt] (1,1) -- (4,4);
	\draw[white, line width = 3pt] (4,4) -- (7,1);
	\draw[white, line width = 3pt] (1,9) -- (4,6);
	\draw[white, line width = 3pt] (4,6) -- (7,9);
	\draw[black, line width = 1pt] (1,9) -- (4,6);
	\draw[black, line width = 1pt] (4,6) -- (7,9);
	\draw[black, line width = 1pt] (1,1) -- (4,4);
	\draw[black, line width = 1pt] (4,4) -- (7,1);
	\draw[black, dashed, line width = 1pt] (4,4) -- (4,6);
	\foreach \r in {1,...,3}{
		\fill[black] (2 + \r/4 ,3/2 ) circle (1/16);
		\fill[black] (2 + \r/4 ,17/2 ) circle (1/16);
		\fill[black] (5 + \r/4 ,3/2 ) circle (1/16);
		\fill[black] (5 + \r/4 ,17/2 ) circle (1/16);
	}
	\draw[decorate, decoration = {brace, mirror, amplitude = 3 pt}, yshift = -3pt] (1.5,.5) -- (6.5,.5) node [midway,yshift = -7pt] {\footnotesize{k}};
	\end{tikzpicture} \; (1_{2^{\otimes n-1}} \otimes J_{k_{1}} \otimes J_{k_{2}} \otimes \hdots \otimes J_{k_{m}} \otimes 1_{2}).
\end{equation}
Since the $J_{k_{j}}$'s are all idempotents, so is their tensor product which can then be decomposed as a sum of primitive idempotents. Since these primitive idempotents are not uniquely defined, there are many ways of performing this decomposition, see for instance \cite{RegularFusion} for examples on how this can be done; we find
\begin{equation}
\tilde{H}^{(k_1,k_2,\hdots, k_{m})}_{2}[n] \sim \bigoplus_{s=1}^{p} \Lambda_{k_{1},k_{2},\hdots ,k_{m}}^{s} \tilde{H}^{(s)}_{2}[n],
\end{equation}
where the sum is a direct sum of matrices, and the coefficients $\Lambda_{k_{1},k_{2},\hdots ,k_{m}}^{s}$ are the multiplicities of each block, which can be obtained from the fusion category:
\begin{equation}
	k_{1}\otimes k_{2} \otimes \hdots \otimes k_{m} = \sum_{s= 1}^{p} \Lambda_{k_{1},k_{2},\hdots ,k_{m}}^{s} s.
\end{equation}
In other words, the fusion of defects reproduce the tensor product of the underlying fusion category.

\subsection{Crossed channel}
In the crossed channel, defects simply appear as polynomials in the hoop operators, which are central elements of the affine Temperley-Lieb algebra. As such, adding multiple defects is quite straightforward: one simply multiplies the central elements representing them. A basis for the space of ``under" $s$-type defects in the crossed channel is simply
\begin{equation}\label{eq:higherspincrossChannel}
	Y^{(k)}_{a}[n] \equiv \mathsf{C}_{k-1}(Y_{a}[n]/2) \;, \qquad k = 0, \hdots p-1,
\end{equation}
where $\mathsf{C}_{k}(x)$ is the $k$th order Chebyshev polynomial of the second kind. Using the properties of these polynomials together with the eigenvalues of $Y_{2}[n]$ obtained in \eqref{eq:hoop.eigenvalues}, one can show that
\begin{equation}
	\prod_{j=1}^{m} Y^{(k_{j})}_{2}[n] = \sum_{s=1}^{p}\Lambda_{k_{1},k_{2},\hdots ,k_{m}}^{s} Y^{(s)}_{2}[n].
\end{equation}
Note that the action of these operators is particularly simple when expressed in terms of anyons; using our previous notation:
\begin{align}\label{eq:defectaction}
	Y^{(k)}_{2}U_{r}[n] \; & = \;
	\begin{tikzpicture}[baseline = {(current bounding box.center)}, every node/.style={scale = 1}, scale = 2/3]
	\foreach \r in {2,3,4,5,6,7,8,9}{
	\node (a\r) at (\r,3) {$\footnotesize{2}$};
	}
	\foreach \r in {5,9,13,17}{
	\draw[black, dashed, line width = 1pt] (\r/2,2) -- (\r/2,1);
	}
	\draw[black, line width = 1pt] (a2) -- (5/2,2);
	\draw[black, line width = 1pt] (a3) -- (5/2,2);
	\draw[black, line width = 1pt] (a4) -- (9/2,2);
	\draw[black, line width = 1pt] (a5) -- (9/2,2);
	\draw[black, line width = 1pt] (a7) -- (13/2,2);
	\draw[black, line width = 1pt] (a6) -- (13/2,2);
	\draw[black, line width = 1pt] (a9) -- (17/2,2);
	\draw[black, line width = 1pt] (a8) -- (17/2,2);
	\node (a0) at (1,1) {$\footnotesize{r}$};
	\node (a11) at (10,1) {$\footnotesize{r}$};
	\draw[black, line width = 1pt] (a0) -- (9/2,1) -- (11/2,.5) -- (6.5,1) -- (a11);
	\node (b0) at (1,-1) {$\footnotesize{k}$};
	\node (b11) at (10,-1) {$\footnotesize{k}$};
	\draw[black, line width = 1pt] (b0) -- (9/2,-1) -- (11/2,-.5) -- (6.5,-1) -- (b11);
	\draw[black, dashed, line width = 1pt] (11/2,.5) -- (11/2,-.5);
	\end{tikzpicture}  \; \notag \\
	& = \sum_{\underset{\text{step } =2}{s = 1+|r-k|}}^{\min( r+ k -1, 2 p - (r+k)+1)} U_{s}[n].
\end{align}

\subsection{Mixed channels}\label{sec:mixedmerge}
We now tackle the cases where one adds multiple defects in different channels; as before, defects in the direct channel are obtained by deforming the Hamiltonian by replacing certain generators, and defects in the crossed channel appear as central elements. In order to understand how these defects interact with each other, we must therefore understand how the eigenvalues of these central elements are modified by the addition of these deformed generators.

Since the defects $ Y_{2}[n], \bar{Y}_{2}[n] $ are central, their eigenvalues can be used to (partially) classify the various $\atl{n}$-modules appearing in the Hilbert space of the RSOS models, with or without defects. We thus denote $\mathcal{H}_{x,y}[n]$, $x,y = 1, \hdots, p $ the sector on which
\begin{equation}\label{eq:def.twistedsectors}
	Y_{2}[n]_{\mathcal{H}_{x,y}[n]} \sim 2\cos\big(\frac{x \pi}{p+1}\big)1_{\mathcal{H}_{x,y}[n]}, \; \quad \bar{Y}_{2}[n]_{\mathcal{H}_{x,y}[n]} \sim 2\cos\big(\frac{y \pi}{p+1}\big)1_{\mathcal{H}_{x,y}[n]} \;.
\end{equation}
One can show that each of these sectors is a simple\footnote{In a more standard notation \cite{GraLeh}, these modules are $\mathcal{H}_{x,y}[n] \sim \mathsf{L}_{|x-y|,\q^{x}}$.} $\atl{n}$-module, and that these are the only ones appearing in the RSOS models, with or without defects. In this notation, the Hilbert space of the periodic $A_{p}$ RSOS model is simply
\begin{equation}\label{eq:SectorPeriodic}
	\mathcal{H}_{\text{period.}}[n] \sim \bigoplus_{x=1}^{p} \mathcal{H}_{x,x}[n].
\end{equation}
Similarly, the same model now carrying a string of ''under" defects $k_{1},k_{2}, \hdots, k_{m}$ would now be
\begin{equation}
	\mathcal{H}^{(k_{1},k_{2},\hdots, k_{m})}[n] \sim \bigoplus_{x=1}^{p} \bigoplus_{y=1}^{p}\Lambda^{y}_{x, k_{1},k_{2},\hdots, k_{m}}\mathcal{H}_{y,x}[n],
\end{equation}
while if those defects were ``over", the decomposition would be
\begin{equation}
	\mathcal{H}^{(k_{1},k_{2},\hdots, k_{m})}[n] \sim \bigoplus_{x=1}^{p} \bigoplus_{y=1}^{p}\Lambda^{y}_{x, k_{1},k_{2},\hdots, k_{m}}\mathcal{H}_{x,y}[n].
\end{equation}
Note in particular that the two types of defects in the same channel are \emph{blind} to each other's presence; this is not the case when they are in different channels.

\section{Quantum inverse scattering and the continuum limit}\label{sec:topdefect.analytics}

While the defects of type $s$ could be studied extensively through exact algebraic methods, the defects of type $r$ require more complex analytical tools; in this section we describe the result of such analysis for certain specific models. For lattice models defined through a transfer matrix $T$, the conformal dimension of the fields appearing in the continuum limit, assuming the models tend to conformal field theories, can be extracted from the spectrum of $T$ through the so-called quantum inverse scattering. We selected three physically relevant models, constructed a transfer matrix representing each of them with the defects we have introduced in this work, and then used this method to obtain their partition function in the continuum limit. We also show the modular $S$-transform of each of those functions; this transformation should correspond to interchanging the crossed and the direct channel, and, as expected, this is indeed what we see in our calculations, at least for defects of type $s$.

We start with some generalities which we used to study these defects, then move on to the specific models we have chosen.
\subsection{Generalities: transfer matrices and spectral curves}
In general, the transfer matrix of the periodic $\mathsf{A}_{p}$ RSOS model (without any defects) can be written as
\begin{equation}
	\mathsf{T}_{n}[u] \equiv \;
	 \begin{tikzpicture}[baseline = {(current bounding box.center)}, scale = 1/2]
	 \draw[black, line width = 1pt] (0,2) -- (6,2);
	\foreach \r in {1,2,4,5}{
		\draw[black, line width = 1pt] (\r,1) -- (\r,3);
		\draw[black, line width = 1pt] (\r,2) circle (1/8);
		\fill[gray] (\r,2) circle (1/8);
		\node[anchor = south,style={scale = 2/3}] at (\r,3) {$2$};
		\node[anchor = north,style={scale = 2/3}] at (\r,1) {$2$};
	}
	\node[anchor = south,style={scale = 2/3}] at (0.5,2) {$2$};
	\node[anchor = south] at (3,2) {$\hdots $};
	\node[anchor = north] at (3,2) {$\hdots $};
	\draw[black, dashed, line width = 1pt] (0,1) -- (0,3);
	\draw[black, dashed, line width = 1pt] (6,1) -- (6,3);
	\end{tikzpicture} \;,
\end{equation}
where the gray vertices (evaluated at $x = u $) were introduced in equation \eqref{eq:def.graydot}. Note that this is precisely the defect operator $\mathbb{Y}_{1}(u)$, so $\mathsf{T}_n[u]$  coincides with the $r$-type defect operator of width $k=1$. The same model, now carrying a $r$-type defect of width $k$, is instead described by the transfer matrix
\begin{align}
	\mathsf{T}^{(k)}_{n}[u;w] & \equiv   \frac{u}{w} (-\q)^{-1/2} \;
		 \begin{tikzpicture}[baseline = {(current bounding box.center)}, scale = 1/2]
	\foreach \r in {5,7}{
		\draw[black, line width = 1pt] (\r,1) -- (\r,3);
		\node[anchor = south,style={scale = 2/3}] at (\r,3) {$2$};
		\node[anchor = north,style={scale = 2/3}] at (\r,1) {$2$};
	}
	 \draw[white, line width = 3pt] (0,2) -- (12,2);
	 \draw[black, line width = 1pt] (0,2) -- (12,2);
	\foreach \r in {1,2,4,8,10,11}{
		\draw[black, line width = 1pt] (\r,1) -- (\r,3);
		\draw[black, line width = 1pt] (\r,2) circle (1/8);
		\fill[gray] (\r,2) circle (1/8);
		\node[anchor = south,style={scale = 2/3}] at (\r,3) {$2$};
		\node[anchor = north,style={scale = 2/3}] at (\r,1) {$2$};
	}
	\fill[white] (4.75,2.25) -- (7.25,2.25) -- (7.25,2.75) -- (4.75,2.75) -- (4.75,2.25);
	\draw[black, line width = 1pt] (4.75,2.25) -- (7.25,2.25) -- (7.25,2.75) -- (4.75,2.75) -- (4.75,2.25);
	\fill[white] (4.75,1.75) -- (7.25,1.75) -- (7.25,1.25) -- (4.75,1.25) -- (4.75,1.75);
	\draw[black, line width = 1pt] (4.75,1.75) -- (7.25,1.75) -- (7.25,1.25) -- (4.75,1.25) -- (4.75,1.75);
	\node[style = {scale = 2/3}] at (6,2.5) {$k$};
	\node[style = {scale = 2/3}] at (6,1.5) {$k$};
	\node[anchor = south,style={scale = 2/3}] at (0.5,2) {$2$};
	\node[anchor = south] at (3,2) {$\hdots $};
	\node[anchor = north] at (3,2) {$\hdots $};
	\node[anchor = south] at (9,2) {$\hdots $};
	\node[anchor = north] at (9,2) {$\hdots $};
	\node[anchor = south] at (6,.5) {$\hdots $};
	\node[anchor = north] at (6,3.5) {$\hdots $};
	\draw[black, dashed, line width = 1pt] (0,1) -- (0,3);
	\draw[black, dashed, line width = 1pt] (12,1) -- (12,3);
	\end{tikzpicture} \notag \\
	& \; - \frac{w}{u} (-\q)^{1/2} \;
	\begin{tikzpicture}[baseline = {(current bounding box.center)}, scale = 1/2]
	 \draw[black, line width = 1pt] (0,2) -- (12,2);
	\foreach \r in {1,2,4,8,10,11}{
		\draw[black, line width = 1pt] (\r,1) -- (\r,3);
		\draw[black, line width = 1pt] (\r,2) circle (1/8);
		\fill[gray] (\r,2) circle (1/8);
		\node[anchor = south,style={scale = 2/3}] at (\r,3) {$2$};
		\node[anchor = north,style={scale = 2/3}] at (\r,1) {$2$};
	}
	\foreach \r in {5,7}{
		\draw[white, line width = 3pt] (\r,1) -- (\r,3);
		\draw[black, line width = 1pt] (\r,1) -- (\r,3);
		\node[anchor = south,style={scale = 2/3}] at (\r,3) {$2$};
		\node[anchor = north,style={scale = 2/3}] at (\r,1) {$2$};
	}
	\fill[white] (4.75,2.25) -- (7.25,2.25) -- (7.25,2.75) -- (4.75,2.75) -- (4.75,2.25);
	\draw[black, line width = 1pt] (4.75,2.25) -- (7.25,2.25) -- (7.25,2.75) -- (4.75,2.75) -- (4.75,2.25);
	\fill[white] (4.75,1.75) -- (7.25,1.75) -- (7.25,1.25) -- (4.75,1.25) -- (4.75,1.75);
	\draw[black, line width = 1pt] (4.75,1.75) -- (7.25,1.75) -- (7.25,1.25) -- (4.75,1.25) -- (4.75,1.75);
	\node[style = {scale = 2/3}] at (6,2.5) {$k$};
	\node[style = {scale = 2/3}] at (6,1.5) {$k$};
	\node[anchor = south,style={scale = 2/3}] at (0.5,2) {$2$};
	\node[anchor = south] at (3,2) {$\hdots $};
	\node[anchor = north] at (3,2) {$\hdots $};
	\node[anchor = south] at (9,2) {$\hdots $};
	\node[anchor = north] at (9,2) {$\hdots $};
	\node[anchor = south] at (6,.5) {$\hdots $};
	\node[anchor = north] at (6,3.5) {$\hdots $};
	\draw[black, dashed, line width = 1pt] (0,1) -- (0,3);
	\draw[black, dashed, line width = 1pt] (12,1) -- (12,3);
	\end{tikzpicture}\;,
\end{align}
where the crossings were defined in equation \eqref{eq:def.black} and \eqref{eq:def.white}; the white boxes are Jones-Wenzl idempotents on $k$ sites, and $\q \equiv e^{i \frac{\pi}{p+1}}$.

More explicitly (after a re-normalisation and a similarity transformation) we find the matrix elements
\begin{equation}
	(\mathsf{T}_{n}^{(k)}[u;w])_{h}^{h'} = \left[\prod_{i=1}^{n} W^{(1)} \! \! \left(\begin{smallmatrix}h_{i}' & h_{i+1}' \\ h_i & h_{i+1} \end{smallmatrix}\Big| u  \right)\right] W^{(k)} \! \! \left(\begin{smallmatrix}h_{n+1}' & h_{1}' \\ h_{n+1} & h_{1} \end{smallmatrix}\Big| \frac{u}{w}  \right),
\end{equation}
where the Boltzmann weights are
\begin{align}\label{IRFweightsS}
W^{(k)}\! \! \left(\begin{smallmatrix}h & h' \\ h \pm_{a}1 & h' \pm_{b} 1 \end{smallmatrix}\Big| u  \right)= \alpha_{\pm_{a}, \pm_{b}} \frac{\left[\frac{\pm_{a}h \pm_{b} h' - k }{2} \right]_{\q}}{\left[h' \right]_{\q}} \big( \frac{\q^\frac{\pm_{a}h \mp_{b} h'}{2}}{u} - \frac{u}{\q^\frac{\pm_{a}h \mp_{b} h'}{2}} \big) \;,
\end{align}
where the two plus/minus signs (denoted $\pm_{a}, \pm_{b}$) vary independently, and
\begin{align}
	\alpha_{+,+} = - (-1)^{\frac{ h' - h - k}{2}} \;, \qquad & \alpha_{-,-} =  (-1)^{\frac{-h' + h - k}{2}} \;, \notag \\
	\alpha_{-,+} = 1, \qquad & \alpha_{+,-} = (-1)^{k} \;,
\end{align}
and the definition of the RSOS models further imposes the constraints
\begin{equation}
	h,h' \in \lbrace 1,2,\hdots, p \rbrace\;, \qquad |h - h'| \in \lbrace -k, 2-k, \hdots, k-2, k \rbrace,	
\end{equation}
\begin{equation}
	h+h' \in \lbrace k+2, k+4, \hdots, 2(p+1)- (k+2) \rbrace \;.
\end{equation}
Applying the Bethe ansatz \cite{BazRes}, we find $p$ spectral curves for the eigenvalues of this matrix
\begin{multline}
\Lambda_{\ell}(u)= \underbrace{\q^{\ell}{\left( \frac{\q^{\frac{1}{2}}}{u} - \frac{u}{\q^{\frac{1}{2}}}\right)}^{n}\left( \frac{\q^{\frac{k}{2}} w}{u} - \frac{u}{\q^{\frac{k}{2}} w}\right) \prod_{j=1}^{\frac{n+k}{2}} \frac{\frac{\q u}{u_j}-\frac{u_j}{\q u}}{\frac{ u}{u_j}-\frac{u_j}{ u}}}_{\lambda_1(u)}+\\
\underbrace{\q^{-\ell}{\left( \frac{\q^{-\frac{1}{2}}}{u} - \frac{u}{\q^{-\frac{1}{2}}}\right)}^{n}\left( \frac{\q^{-\frac{k}{2}}w }{u} - \frac{u}{\q^{-\frac{k}{2}} w}\right) \prod_{j=1}^{\frac{n+k}{2}} \frac{\frac{ u}{\q u_j}-\frac{\q u_j}{ u}}{\frac{ u}{u_j}-\frac{u_j}{ u}}}_{\lambda_2(u)}\;,
\end{multline}
where $\ell = 1, \hdots, p$, and $\lbrace u_{j} \rbrace$ are the \emph{Bethe roots} of the system ; these are functions of $\q, w,$ and $k$, determined by the \emph{Bethe equations}, a collection of non-linear equations ensuring that the various poles in $\Lambda_{\ell}(u)$ are removable. Using standard techniques, and assuming that the continuum limit of this model is described by a conformal field theory, one can use these expressions to obtain the conformal dimension of the primary fields describing it. The results can be characterized by some general aspects of the distribution of Bethe roots, like the  asymptotic limits of the auxiliary function $\frac{\lambda_{1}(u)}{\lambda_{2}(u)} $, the number of eigenvalue zeroes, and the presence of \emph{strings}, families of roots with the same real part and equally spaced imaginary parts. For more details on how this is done and the explicit arguments leading to the results of this section, see \cite{nextpaper}.

If we let $w \rightarrow 0$ or $w \rightarrow \infty$ we recover the $s$-type defects. In this case, a subset of at most $k$ Bethe roots $u_j$ go to $0,\infty$ as well, thereby giving rise to  modified spectral curves.
\begin{equation}
\Lambda_{\ell,t}^{\pm}(u)= \underbrace{\q^{\ell\pm t}{\left( \frac{\q^{\frac{1}{2}}}{u} - \frac{u}{\q^{\frac{1}{2}}}\right)}^{n} \prod_{j=1}^{\frac{n}{2}+t} \frac{\frac{\q u}{u_j}-\frac{u_j}{\q u}}{\frac{ u}{u_j}-\frac{u_j}{ u}}}_{\lambda_1(u)}+\\
\underbrace{\q^{-(\ell\pm t)}{\left( \frac{\q^{-\frac{1}{2}}}{u} - \frac{u}{\q^{-\frac{1}{2}}}\right)}^{n} \prod_{j=1}^{\frac{n}{2}+t} \frac{\frac{ u}{\q u_j}-\frac{\q u_j}{ u}}{\frac{ u}{u_j}-\frac{u_j}{ u}}}_{\lambda_2(u)}\;,
\end{equation}
where the plus sign corresponds to $w \rightarrow \infty$, the minus sign corresponds to $w \rightarrow 0$ and $t \in \left\{-k/2,-k/2+1,\ldots,k/2\right\}$ describes the number of Bethe roots $\frac{k}{2}-t$ going to $0$ or $\infty$. Because of the symmetry of the possible values of $\ell$ and $t$, one gets the same spectrum for both $s$-type Hamiltonians. Nevertheless, a specific energy level, parameterized in terms of Bethe roots, often leads to different levels depending if one moves adiabatically from $w=1$ to $w=0$ or from $w=1$ to $w=\infty$.


\subsection{$A_{3}$}
As a model on a finite lattice, the periodic $A_{3}$ RSOS model is equivalent to a direct sum of two Ising models: one with periodic boundary conditions, and one with anti-periodic ones. It follows that its continuum limit is described by (two copies of) the Virasoro minimal model $M(4,3)$. Inserting our various defects in the direct channel, we obtain the following three twisted partition functions\footnote{Here, and in the following, we suppress the global factor of two inherent to these models, in order to lighten the notation.}:
\begin{align*}
		Z_{1,1} & \equiv  |\chi_{1,1}|^{2} + |\chi_{1,2}|^{2} + |\chi_{1,3}|^{2}, \\
		Z_{1,2} & \equiv  (\chi_{1,1} + \chi_{1,3})\bar{\chi}_{1,2} +  \chi_{1,2}(\bar{\chi}_{1,1} +\bar{\chi}_{1,3}),\\
		Z_{1,3} & \equiv  (\chi_{1,1} \bar{\chi}_{1,3} + \bar{\chi}_{1,1}\chi_{1,3}) +  \chi_{1,2}\bar{\chi}_{1,2},
\end{align*}
where the indices in $Z_{r,1 + s}$ measures the width of the added defects of type $r$ and $s$, respectively, and $\chi_{r,s}$ is the Virasoro character with Kac label $(r,s)$. Taking the modular $S$ transform of these partition functions, we find:
\begin{align*}
		S(Z_{1,1}) & =  |\chi_{1,1}|^{2} + |\chi_{1,2}|^{2} + |\chi_{1,3}|^{2}, \\
		S(Z_{1,2}) & = [2]_{\q}|\chi_{1,1}|^{2} +  [2]_{\q^{2}}|\chi_{1,2}|^{2} + [2]_{\q^{3}}|\chi_{1,3}|^{2},\\
		S(Z_{1,3}) & = [3]_{\q}|\chi_{1,1}|^{2} +  [3]_{\q^{2}}|\chi_{1,2}|^{2} + [3]_{\q^{3}}|\chi_{1,3}|^{2},
\end{align*}
where $[k]_{q} \equiv \frac{\q^{k}-\q^{-k}}{\q-\q^{-1}}$ are $\q$-numbers, with $\q \equiv e^{\frac{i \pi}{4}}$. One recognizes that the weights appearing in these expressions are precisely the eigenvalues of the defect operators acting in the crossed channel (see equations \eqref{eq:higherspincrossChannel}, \eqref{eq:def.twistedsectors}, and \eqref{eq:SectorPeriodic}); this suggests that the defects in the two channels are indeed related by a modular transformation.

\subsection{$A_{4} $}

The periodic $A_{4}$ RSOS model is described, in the continuum limit, by two copies of the Virasoro minimal model $M(5,4)$; on a finite lattice, it contains, but is not equivalent to, the periodic Fibonacci anyon chain of even length (see section \ref{sec:fibon}). Inserting our various defects in the direct channel, we obtain the following six twisted partition functions:
\begin{align*}
		Z_{1,1} & \equiv   |\chi_{1,1}|^{2} +  |\chi_{3,1}|^{2} + |\chi_{2,2}|^{2} + |\chi_{1,3}|^{2} + |\chi_{3,3}|^{2} + |\chi_{2,4}|^{2},\\
		Z_{1,2}  & \equiv ( (\chi_{1,1} + \chi_{1,3})\bar{\chi}_{3,3} + \chi_{2,2}\bar{\chi}_{2,4} + \chi_{1,3}\bar{\chi}_{3,1} + c.c.) + |\chi_{2,2}|^{2} ,\\
		Z_{1,3} & \equiv (\chi_{1,1}\bar{\chi}_{1,3} + \chi_{2,4}\bar{\chi}_{2,2} + \chi_{3,1}\bar{\chi}_{3,3} + c.c.) + |\chi_{3,3}|^{2} + |\chi_{2,2}|^{2} + |\chi_{1,3}|^{2},\\
		Z_{1,4} & \equiv (\chi_{1,1}\bar{\chi}_{3,1} + \chi_{1,3}\bar{\chi}_{3,3} + c.c.) + |\chi_{2,4}|^{2} + |\chi_{2,2}|^{2},\\
		Z_{2,1} & \equiv (\chi_{1,1} + \chi_{3,1})\bar{\chi}_{2,4} +(\chi_{3,3} + \chi_{1,3})\bar{\chi}_{2,2} + c.c. ,\\
		Z_{2,2} & \equiv (\chi_{1,1} + \chi_{3,1})\bar{\chi}_{2,2} +(\chi_{1,3} + \chi_{3,3})(\bar{\chi}_{2,1}+\bar{\chi}_{2,2}) + c.c. ,
	\end{align*}
where $c.c.$ stands for the complex conjugate. Taking the modular $S$ transform of these partition functions, we find:
\begin{align*}
		S(Z_{1,1}) & =  |\chi_{1,1}|^{2} +  |\chi_{3,1}|^{2} + |\chi_{2,2}|^{2} + |\chi_{1,3}|^{2} + |\chi_{3,3}|^{2} + |\chi_{2,4}|^{2},\\
		S(Z_{1,2})& = [2]_{q}(|\chi_{1,1}|^{2} +  |\chi_{3,1}|^{2}) +  [2]_{q^{2}}|\chi_{2,2}|^{2} +  [2]_{q^{3}}(|\chi_{1,3}|^{2} + |\chi_{3,3}|^{2}) +[2]_{q^{4}} |\chi_{2,4}|^{2},\\
		S(Z_{1,3})& = [3]_{q}(|\chi_{1,1}|^{2} +  |\chi_{3,1}|^{2}) +  [3]_{q^{2}}|\chi_{2,2}|^{2} +  [3]_{q^{3}}(|\chi_{1,3}|^{2} + |\chi_{3,3}|^{2}) +[3]_{q^{4}} |\chi_{2,4}|^{2},\\
		S(Z_{1,4})& =  [4]_{q}(|\chi_{1,1}|^{2} +  |\chi_{3,1}|^{2}) +  [4]_{q^{2}}|\chi_{2,2}|^{2} +  [4]_{q^{3}}(|\chi_{1,3}|^{2} + |\chi_{3,3}|^{2}) +[4]_{q^{4}} |\chi_{2,4}|^{2},\\
		S(Z_{2,1}) & = ([2]_{\Q}|\chi_{1,1}|^{2} +  [2]_{\Q^{3}}|\chi_{3,1}|^{2}) + [2]_{\Q^{2}}|\chi_{2,2}|^{2} \\
		& \quad + ([2]_{\Q}|\chi_{1,3}|^{2} + [2]_{\Q^{3}}|\chi_{3,3}|^{2}) + [2]_{\Q^{2}}|\chi_{2,4}|^{2},\\
		S(Z_{2,2}) & = [2]_{q}([2]_{\Q}|\chi_{1,1}|^{2} +  [2]_{\Q^{3}}|\chi_{3,1}|^{2}) + [2]_{q^{2}}[2]_{\Q^{2}}|\chi_{2,2}|^{2} + [2]_{q^{3}}([2]_{\Q}|\chi_{1,3}|^{2}\\
		& \quad + [2]_{\Q^{3}}|\chi_{3,3}|^{2}) + [2]_{q^{4}}[2]_{\Q^{2}}|\chi_{2,4}|^{2},
	\end{align*}
where $\q = e^{\frac{i \pi}{5}}$ and $\Q = e^{\frac{i \pi}{4}} $. Again we note that the coefficients appearing in the $S$-transform of the partition functions only carrying $s$-type defects (namely $S(Z_{1,1+s})$) are indeed the eigenvalues of the defect operators in the crossed channel (see equations \eqref{eq:higherspincrossChannel}, \eqref{eq:def.twistedsectors}, and \eqref{eq:SectorPeriodic}).

\subsection{The general case}

We find in general that the defect of $(r,s)$-type with $(k,k')$ lines corresponds to the $(r,s)=(\hbox{max }(1,k),k'+1)$ defect in the diagonal minimal model (\ref{CFTz}) (the condition on $r$ stems from the fact that having an $r$-type defect with $k=1$ or no defect amounts to no defect in the continuum limit.  We note in particular that the ``over'' and ``under'' versions of the $s$-type defect correspond to the same object in the continuum limit. In more general cases (that is, not necessarily diagonal), we think over and under correspond to chiral and anti-chiral versions of the defect, giving rise to partition functions related by conjugation $q\to \bar{q}$, where $\q $ (resp. $\bar{\q}$) are the modular parameters for the left and right moving sector. This difference is not seen for the anyonic chains because they correspond to diagonal theories, so fusion with $\phi_{1,s}$ and $\bar{\phi}_{1,s}$ give the same result. Recall from \eqref{CFTz} that the model without any defect has the partition function
\begin{equation}
		Z \sim \sum_{r = 1}^{p-1}\sum_{s=1}^{p} |\chi_{r,s}|^{2},
	\end{equation}
	where $\chi_{r,s}$ is the character of the irreducible Virasoro modules with Kac label $(r,s)$. Inserting the chiral defect yields the partition function
	\begin{equation}
		Z_{1,k} \sim \sum_{r = 1}^{p-1}\sum_{s=1}^{p} (\chi_{r,s} \times_{f} \chi_{1,k}) \bar{\chi}_{r,s},
	\end{equation}
	where it is understood that the fusion product $\times_{f}$ of characters stands for the character of the Virasoro fusion product of the corresponding irreducible modules \cite{YellowBible}. Similarly, inserting the anti-chiral defect yields
	\begin{equation}
		Z_{1,k^{*}} \sim \sum_{r = 1}^{p-1}\sum_{s=1}^{p} \chi_{r,s}(\bar{\chi}_{r,s}\times_{f} \bar{\chi}_{1,k}).
	\end{equation}
This, meanwhile, is not true for non-diagonal theories. We depart from our focus on $A_p$ anyonic chains to illustrate this point now.

\subsubsection{$A_{5}$ -- Three-state Potts models}

The $A_{5}$ RSOS model corresponds to the Virasoro minimal model $M(6,5)$ which admits two distinct modular invariant partition functions \cite{YellowBible} ; we focus here on the non-diagonal one corresponding to the three-state Potts model. Note that it is usually described by the periodic RSOS model of type $\mathsf{D}_{4}$, which is unfortunately outside of the scope of this work. However, one can show that the periodic RSOS models of type $\mathsf{D}$ are equivalent to models of type $\mathsf{A}$ with twisted boundary conditions; in this case it corresponds to the twisted model with sectors (see \eqref{eq:def.twistedsectors})
\begin{equation}
	\mathcal{H}_{\rm Potts}[n] \sim \mathcal{H}_{1,1}[n] + \mathcal{H}_{3,3}[n] + \mathcal{H}_{1,5}[n].
\end{equation}
Applying our defects procedure yields ten distinct twisted partition functions:
\begin{align*}
	Z_{1,1} & \equiv \sum_{r=1}^{4} \big( |\chi_{r,1}|^{2} + |\chi_{r,3}|^{2} + \chi_{r,1}\bar{\chi}_{r,5}\big) = \sum_{r=1,2}\big( |\chi_{r,1} + \chi_{r,5}|^{2} + 2 |\chi_{r,3}|^{2} \big),\\
	Z_{1,2} & \equiv (\chi_{2,2} + \chi_{3,2})(\bar{\chi}_{2,1} + \bar{\chi}_{3,1} + 2\bar{\chi}_{3,3}) + (\chi_{1,2} + \chi_{4,2})(\bar{\chi}_{1,1} + \bar{\chi}_{4,1} + 2\bar{\chi}_{4,3}), \\
	Z_{1,2^{*}} & \equiv \bar{Z}_{1,2},\\
	Z_{1,2^{*} 2} & \equiv 3\big(|\chi_{2,2} + \chi_{3,2}|^{2} + |\chi_{4,2} + \chi_{4,4}|^{2} \big),\\
	Z_{1,3} & \equiv
2\big( \chi_{3,3}(\bar{\chi}_{2,1} + \bar{\chi}_{3,1}) + \chi_{4,3}(\bar{\chi}_{1,1} + \bar{\chi}_{4,1}) + c.c.\big) + 2 |\chi_{3,3}|^{2} + 2 |\chi_{4,3}|^{2} ,\\
	Z_{2,1}& \equiv \big( |\chi_{1,1} + \chi_{2,1} +\chi_{3,1} + \chi_{4,1}|^{2} - |\chi_{1,1} + \chi_{4,1}|^{2}\big) + 2 \big( |\chi_{3,3} + \chi_{4,3}|^{2} - |\chi_{4,3}|^{2}\big),\\
	Z_{2,2} & \equiv \big(\chi_{4,2} + \chi_{4,4} + \chi_{2,2} + \chi_{3,2} \big)\big(\bar{\chi}_{2,1} + \bar{\chi}_{3,1} + 2\bar{\chi}_{3,3} \big) \\
	& \quad + \big( \chi_{2,2} + \chi_{3,2}\big)\big( \bar{\chi}_{1,1} + \bar{\chi}_{4,1} + 2\bar{\chi}_{4,3} \big),\\
	Z_{2,2^{*}}& \equiv \bar{Z}_{2,2},\\
	Z_{2,2^{*}2} & \equiv 3 \big(|\chi_{1,2} + \chi_{2,2} +\chi_{3,2} + \chi_{4,2}|^{2} - |\chi_{1,2} + \chi_{4,2}|^{2} \big),\\
	Z_{2,3} & \equiv 2\big( \big(\chi_{2,1} + \chi_{3,1}\big)\big(\bar{\chi}_{3,3} + \bar{\chi}_{4,3}\big)+\big(\chi_{1,1} + \chi_{4,1} + \chi_{4,3}\big)\bar{\chi}_{3,3} + c.c. \big) + 2|\chi_{3,3}|^{2},
\end{align*}
where the notation $k^{*}$ stands for the insertion of an anti-chiral defect.

The partition function $Z_{2,2*2}$ for instance is obtained in the direct channel by fusion with $\phi_{21}$, $\bar{\phi}_{12}$ and $\phi_{12}$.
\
Taking the modular transform of these characters, we find
\begin{align*}
	S(Z_{1,1}) & = \sum_{r=1}^{4} \big( |\chi_{r,1}|^{2} + |\chi_{r,3}|^{2} + \chi_{r,1}\bar{\chi}_{r,5}\big),\\
	S(Z_{1,2}) & = \sum_{r=1}^{4} (-1)^{r+1}\big( [2]_{\q}|\chi_{r,1}|^{2} + [2]_{\q^{3}}|\chi_{r,3}|^{2} + [2]_{\q}\chi_{r,1}\bar{\chi}_{r,5}\big),\\
	S(Z_{1,2^{*}}) & = \sum_{r=1}^{4} (-1)^{r+1}\big( [2]_{\q}|\chi_{r,1}|^{2} + [2]_{\q^{3}}|\chi_{r,3}|^{2} + [2]_{\q^{5}}\chi_{r,1}\bar{\chi}_{r,5}\big),\\
	S(Z_{1,2^{*}2}) & = \sum_{r=1}^{4} \big( [2]_{\q}[2]_{\q}|\chi_{r,1}|^{2} + [2]_{\q^{3}}[2]_{\q^{3}}|\chi_{r,3}|^{2} + [2]_{\q}[2]_{\q^{5}}\chi_{r,1}\bar{\chi}_{r,5}\big),\\
	S(Z_{1,3}) & = \sum_{r=1}^{4} \big( [3]_{\q}|\chi_{r,1}|^{2} + [3]_{\q^{3}}|\chi_{r,3}|^{2} + [3]_{\q^{5}}\chi_{r,1}\bar{\chi}_{r,5}\big),
\end{align*}
and
\begin{align*}
	S(Z_{2,1}) & = \sum_{r=1}^{4} (-1)^{r+1}[2]_{\Q^{r}}\big( |\chi_{r,1}|^{2} + |\chi_{r,3}|^{2} + \chi_{r,1}\bar{\chi}_{r,5}\big),\\
	S(Z_{2,2}) & = \sum_{r=1}^{4}[2]_{\Q^{r}}\big( [2]_{\q}|\chi_{r,1}|^{2} + [2]_{\q^{3}}|\chi_{r,3}|^{2} + [2]_{\q}\chi_{r,1}\bar{\chi}_{r,5}\big),\\
	S(Z_{2,2^{*}}) & = \sum_{r=1}^{4}[2]_{\Q^{r}}\big( [2]_{\q}|\chi_{r,1}|^{2} + [2]_{\q^{3}}|\chi_{r,3}|^{2} + [2]_{\q^{5}}\chi_{r,1}\bar{\chi}_{r,5}\big),\\
	S(Z_{2,2^{*}2}) & = \sum_{r=1}^{4}(-1)^{r+1}[2]_{\Q^{r}}\big( ([2]_{\q})^{2}|\chi_{r,1}|^{2} + ([2]_{\q^{3}})^{2}|\chi_{r,3}|^{2} + [2]_{\q}[2]_{\q^{5}}\chi_{r,1}\bar{\chi}_{r,5}\big),\\
	S(Z_{2,3}) & = \sum_{r=1}^{4}(-1)^{r+1}[2]_{\Q^{r}}\big( [3]_{\q}|\chi_{r,1}|^{2} + [3]_{\q^{3}}|\chi_{r,3}|^{2} + [3]_{\q}\chi_{r,1}\bar{\chi}_{r,5}\big),
\end{align*}
where $\q = e^{\frac{i \pi}{6}}$ and $\Q = e^{\frac{i \pi}{5}}$. Note that in addition to the coefficients appearing in front of each character (which are still the eigenvalues of the crossed channel defect operators) there is an additional sign change; this happens because of our choice of basis for the set of fields, since in this model $\chi_{2,1} \equiv \chi_{3,5} $, but $[2]_{\q} = - [2]_{\q^{5}}$.

\subsection{The physics of defects of $r$-type}

In contrast with the defects of type $s$, the defects of type $r$ are not topological on the lattice.  However,  analytical and numerical calculations shows that their introduction in the lattice partition function has precisely the same impact as the introduction of a topological defects should have in the continuum. This suggests that these are \emph{asymptotically topological}, only commuting with the actual conformal generators, not their lattice version. The existence of such behaviour is not a priori surprising, since the whole richness of conformal symmetry is never fully preserved on a finite lattice. Something a little deeper happens, however. It is often understood that the Temperley-Lieb algebras do not correspond to a lattice version of the bulk conformal algebra $ \mathsf{Vir} \otimes \bar{\mathsf{Vir}}$, but rather to a larger algebra dubbed the \emph{inter-chiral}\footnote{This algebra closely resembles $ \mathsf{Vir} \otimes \bar{\mathsf{Vir}}$ but also contains generators which \emph{mixes} the left and the right Virasoro algebras, hence the name \emph{inter-chiral.}} algebra \cite{Interchiral}. The  defects of type $r$  are topological in the continuum limit, but do not commute with the  full inter-chiral algebra. In contrast, defects of type $s$ are, in a certain sense, more than topological, since they  commute with the full inter-chiral algebra. They also commute with the lattice version of this algebra, which is the Temperley-Lieb algebra, hence appear ``topological'' on the lattice as well.

The difference between defects of type $s$ and defects of type $r$ is actually quite interesting  from the point of view of the renormalization group. Insertion of a defect of type $s$ on the lattice corresponds, in the continuum limit,  to the insertion of a topological defect parametrized by the Kac labels $(1,k)$ . Insertion of a defect of type $r$ is obtained with the same number of lines $k$, and a finite (as opposed to infinite or zero) values of $w$. This corresponds, in the continuum limit, to a {\sl perturbation} of the topological defect carrying Kac labels $(1,k)$
 by a relevant operator which induces a renormalization group flow \cite{KRW}. The flow terminates at an infrared fixed point, which turns out to correspond to the topological defect parametrized by Kac labels $(k-1,1)$. In other words, the lattice model provides for generic $w$, the spectral parameter, a generic defect (not topological, not conformal) that interpolates between a topological defect in the UV and in the IR. Any finite non-zero value of $w$ leads, for large systems, to IR physics, while, for  a zero or infinite value, we see the UV physics.

To be a little more quantitative, the perturbation at finite $w$ corresponds to the conformal perturbation by $\phi_{13}$ or $\bar{\phi}_{13}$ (depending on the sign of $\ln {w\over u}$). The perturbation induces a characteristic energy  scale $T_0$ (analogous to the Kondo temperature in Kondo physics) with, when $w$ approaches $0$ or $\infty$,  $T_0\propto e^{-\gamma {|\ln w|\over \pi}}$. As the size of the systems goes from $n \ll {1\over T_0} $ to $n \gg {1\over T_0}$, a crossover between UV and IR physics takes place.

Note that, while we believe that chiral and anti-chiral defects of type $r$ can be obtained by taking $w>u$ (resp. $w<u$), due to the symmetry of the Kac table and the specific combinations of primary fields appearing in the continuum limit of the RSOS models, it turns out that the insertion of a defect of this type in either sectors produces exactly the same partition function {\sl even for non-diagonal minimal models}, that is
\begin{equation}
	\sum_{r = 1}^{p} (\chi_{r,x} \times_{f} \chi_{\rho,1} )\bar{\chi}_{r,y} = \sum_{r = 1}^{p} \chi_{r,x} (\bar{\chi}_{r,y} \times_{f} \bar{\chi}_{\rho,1} ),
\end{equation}
where the the fusion product of characters stands for the character of the Virasoro fusion product of the corresponding primary fields. In other words, if chiral and anti-chiral defects of type $r$ exists, the RSOS models are entirely blind to any distinctions between the two.

\section{Conclusions and outlook}

Since deformations of the defect lines in our models are generated by the Temperley-Lieb algebra, it is difficult to imagine a defect of type $r$ that would appear topological on the lattice as well.  Indeed,  while in this work we have identified two families of central elements $Y^{(k)}_{2},\bar{Y}^{(k)}_{2}$, and explained how their eigenvalues can be used to distinguished all sectors of these lattice models, it turns out that the whole center of $\atl{n}$ is spanned by polynomials of arbitrary degrees in $Y^{(1)}_{2}$ and $\bar{Y}^{(1)}_{2}$. There is just ``nothing else" that would commute with line deformations in general, and provide other types of defects: a defect that is topological on the lattice necessarily is a combination of defects of type $s$. Note this is true only for the generic case - relevant, e.g., for the $Q$-state Potts model, which is formulated in terms of lines. It is not impossible to imagine there exists an entirely different version of the universality classes in (\ref{CFTz}) that would not involve the Temperley-Lieb algebra, and for which the center of the lattice algebra generating deformations of the defect line would be bigger. We do not really believe this is possible, however. In any case,
like in many similar instances \cite{PS}, we are, for now,  stuck with being able to reproduce only ``half'' of the whole structure of the CFT exactly on the lattice. 

Several interesting points must finally be mentioned. First, since the lattice defects of type $s$ are built entirely using the center  of the affine Temperley-Lieb algebra, they will be topological on the lattice and therefore in the continuum limit for all Temperley-Lieb Hamiltonians, not only the simple one (\ref{Hamil}) discussed here. Interesting cases include for instance the $Z_n$ parafermionic theories occurring at RSOS points of  the antiferromagnetic Potts model \cite{JS},  the critical points in the Majumdar-Ghosh-like Temperley-Lieb models \cite{IJS}, etc. In contrast, we do not know what happens to the defects of type $r$ in these other cases: since they are not topological on the lattice, more work is needed to find out what happens to them in the continuum limit.

Second, the topological symmetry identified in \cite{Feiguinetal,Gilsetal} was in fact known in the field theory community under the name ``residual quantum symmetry'', and first identified in works by Bernard and Leclair \cite{BL}: the construction of the action of the symmetry using 6j-coefficients in that paper is virtually identical to calculations in \cite{Gilsetal}.

Third, what is meant in \cite{Feiguinetal} by ``topological protection'' is that, since the anyonic Hamiltonian has topological symmetry, it commutes with the generators $Y_2,\bar{Y}_2$ of the center of the Temperley-Lieb algebra. This implies that some potential perturbations of the CFT Hamiltonian are absent. The full continuum limit of the affine Temperley-Lieb algebra  is believed to be generated by $ \mathsf{Vir} \otimes \bar{\mathsf{Vir}}$ and by fusion with $(\phi_{12},\bar{\phi}_{12})$\footnote{This is, of course, a perturbative statement, meaning more precisely that if we perturb the critical model (corresponding to the homogeneous Hamiltonian $H\propto \sum e_i$) by local operators belong to ATL, this will translate into perturbations of the CFT by fields within $ \mathsf{Vir} \otimes \bar{\mathsf{Vir}}$, as well as the conformal towers of $(\phi_{1s},\bar{\phi}_{1s})$.}. Hence small perturbations of the Hamiltonian that preserve the topological symmetry can only expand on those fields, which is a much smaller set of potentially relevant perturbations than are present in the whole CFT.

Fourth, the study  of topological defects with the quantum inverse scattering method makes it clear that these objects - or rather, their lattice realizations - are very natural from the point of view of integrable systems: all they boil down to is the insertion of lines of heterogeneities carrying both, in general, a different spin and spectral parameter than those in the bulk. Such construction has a long history in the field of quantum impurity problems \cite{Impurities1,Impurities2,Impurities3,Impurities4}. We repeat, however, that defects \emph{can} be defined in the non-integrable case as well.

\bigskip
\noindent{\bf Acknowledgments:} This work was supported by the ERC Advanced Grant NuQFT. The authors would also like to thank Zhenghan Wang and Romain Couvreur for interesting discussions and comments on this manuscript.

\appendix

\section{Conventions}
\subsection{$F$-moves}\label{sec:fsymbols}
	By definitions, the $F$-matrices encode the associator of the category:
		\begin{equation}
	\begin{tikzpicture}[baseline = {(current bounding box.center)},scale = 2/3]
	\draw[black, line width = 1pt] (1,3) -- (3,1);
	\draw[black, line width = 1pt] (2,3) -- (5/3,7/3);
	\draw[black, line width = 1pt] (3,3) -- (7/3,5/3);
	\node[anchor = south] at (1,3) {$a$};
	\node[anchor = south] at (2,3) {$b$};
	\node[anchor = south] at (3,3) {$c$};
	\node[anchor = north] at (3,1) {$d$};
	\node[anchor = east] at (2,2) {$x$};
	\end{tikzpicture} \; \equiv \sum_{y \in \mathsf{A_{p}}} (\mathsf{F}_{a,b,c}^{d})_{x}^{y} \;
	\begin{tikzpicture}[baseline = {(current bounding box.center)},scale = 2/3]
	\draw[black, line width = 1pt] (1,3) -- (3,1);
	\draw[black, line width = 1pt] (2,3) -- (2.5,2.5);
	\draw[black, line width = 1pt] (3,3) -- (2,2);
	\node[anchor = south] at (1,3) {$a$};
	\node[anchor = south] at (2,3) {$b$};
	\node[anchor = south] at (3,3) {$c$};
	\node[anchor = north] at (3,1) {$d$};
	\node[anchor = north] at (2.5,2.5) {$y$};
	\end{tikzpicture} \; , \qquad
	\begin{tikzpicture}[baseline = {(current bounding box.center)},scale = 2/3]
	\draw[black, line width = 1pt] (1,3) -- (3,1);
	\draw[black, line width = 1pt] (2,3) -- (2.5,2.5);
	\draw[black, line width = 1pt] (3,3) -- (2,2);
	\node[anchor = south] at (1,3) {$a$};
	\node[anchor = south] at (2,3) {$b$};
	\node[anchor = south] at (3,3) {$c$};
	\node[anchor = north] at (3,1) {$d$};
	\node[anchor = north] at (2.5,2.5) {$y$};
	\end{tikzpicture} \; \equiv \sum_{y \in \mathsf{A_{p}}} (\mathsf{F}_{a,b,c}^{d})_{x}^{y} \;
	\begin{tikzpicture}[baseline = {(current bounding box.center)},scale = 2/3]
	\draw[black, line width = 1pt] (1,3) -- (3,1);
	\draw[black, line width = 1pt] (2,3) -- (5/3,7/3);
	\draw[black, line width = 1pt] (3,3) -- (7/3,5/3);
	\node[anchor = south] at (1,3) {$a$};
	\node[anchor = south] at (2,3) {$b$};
	\node[anchor = south] at (3,3) {$c$};
	\node[anchor = north] at (3,1) {$d$};
	\node[anchor = east] at (2,2) {$x$};
	\end{tikzpicture} \;.
	\end{equation}
	Note that these trees are trivial if the morphisms they represents are; for instance
	\begin{equation*}
	\begin{tikzpicture}[baseline = {(current bounding box.center)},scale = 2/3]
	\draw[black, line width = 1pt] (1,3) -- (3,1);
	\draw[black, line width = 1pt] (2,3) -- (5/3,7/3);
	\draw[black, line width = 1pt] (3,3) -- (7/3,5/3);
	\node[anchor = south] at (1,3) {$a$};
	\node[anchor = south] at (2,3) {$1$};
	\node[anchor = south] at (3,3) {$c$};
	\node[anchor = north] at (3,1) {$a$};
	\node[anchor = east] at (2,2) {$x$};
	\end{tikzpicture} \equiv 0 \qquad \text{ unless } x\equiv a ,
	\end{equation*}
	because the top trivalent vertex is trivial if $x\neq a$. By convention, every $F$-symbol corresponding to a trivial tree is set to zero. Furthermore, because the simple objects are all self-dual, we also have
	\begin{equation}
	\begin{tikzpicture}[baseline = {(current bounding box.center)},scale = 2/3]
	\node (A) at (1,1) {$a$};
	\node (B) at (1,3) {$b$};
	\node (C) at (5,3) {$c$};
	\node (D) at (5,1) {$d$};
	\draw[black, line width = 1pt] (A) -- (2,2);
	\draw[black, line width = 1pt] (2,2) -- (B);
	\draw[black, line width = 1pt] (2,2) -- (4,2);
	\draw[black, line width = 1pt] (C) -- (4,2);
	\draw[black, line width = 1pt] (D) -- (4,2);
	\node[anchor = north] at (3,2) {$x$};
	\end{tikzpicture}\; \equiv \sum_{y}(F_{a,b,c}^{d})_{x}^{y} \;
	\begin{tikzpicture}[baseline = {(current bounding box.center)},scale = 2/3]
	\node (A) at (1,1) {$a$};
	\node (B) at (1,5) {$b$};
	\node (C) at (3,5) {$c$};
	\node (D) at (3,1) {$d$};
	\draw[black, line width = 1pt] (A) -- (2,2);
	\draw[black, line width = 1pt] (2,2) -- (D);
	\draw[black, line width = 1pt] (2,2) -- (2,4);
	\draw[black, line width = 1pt] (C) -- (2,4);
	\draw[black, line width = 1pt] (B) -- (2,4);
	\node[anchor = east] at (2,3) {$y$};
	\end{tikzpicture} \;.
	\end{equation}
	Because we assume that the category is strict,
	\begin{equation}\label{eq:crossingsym.fmoves}
	\begin{tikzpicture}[baseline = {(current bounding box.center)},scale = 2/3]
	\node (A) at (1,4) {$a$};
	\node (B) at (0,2) {$a$};
	\node (C) at (2,2) {$1$};
	\node (D) at (1,0) {$a$};
	\draw[black, line width = 1pt] (A) -- (1,3);
	\draw[black, line width = 1pt] (1,3) -- (B);
	\draw[black, line width = 1pt] (1,3) -- (C);
	\draw[black, line width = 1pt] (B) -- (1,1);
	\draw[black, line width = 1pt] (C) -- (1,1);
	\draw[black, line width = 1pt] (1,1) -- (D);
	\end{tikzpicture} \; \equiv \;
	\begin{tikzpicture}[baseline = {(current bounding box.center)},scale = 2/3]
	\node (A) at (1,2) {$a$};
	\node (D) at (1,0) {$a$};
	\draw[black, line width = 1pt] (A) -- (D);
	\end{tikzpicture} \; \equiv \;
	\begin{tikzpicture}[baseline = {(current bounding box.center)},scale = 2/3]
	\node (A) at (1,4) {$a$};
	\node (B) at (0,2) {$1$};
	\node (C) at (2,2) {$a$};
	\node (D) at (1,0) {$a$};
	\draw[black, line width = 1pt] (A) -- (1,3);
	\draw[black, line width = 1pt] (1,3) -- (B);
	\draw[black, line width = 1pt] (1,3) -- (C);
	\draw[black, line width = 1pt] (B) -- (1,1);
	\draw[black, line width = 1pt] (C) -- (1,1);
	\draw[black, line width = 1pt] (1,1) -- (D);
	\end{tikzpicture} \;,
	\end{equation}
	and thus the triangle identity reduces to
	\begin{equation}
		(\mathsf{F}_{a,1,b}^{c})_{a}^{b} \equiv 1,
	\end{equation}
	unless of course $\mathsf{Hom}_{\mathcal{C}_{p}}(a\otimes b ,c) = 0 $. Furthermore if $x = 1$ then equation \eqref{eq:crossingsym.fmoves} reduces to
	\begin{equation}
	\begin{tikzpicture}[baseline = {(current bounding box.center)},scale = 2/3]
	\node (A) at (1,1) {$a$};
	\node (B) at (1,3) {$a$};
	\node (C) at (3,3) {$b$};
	\node (D) at (3,1) {$b$};
	\draw[black, line width = 1pt] (A) -- (B);
	\draw[black, line width = 1pt] (C) -- (D);
	\end{tikzpicture} \; \equiv \;
	\begin{tikzpicture}[baseline = {(current bounding box.center)},scale = 2/3]
	\node (A) at (1,1) {$a$};
	\node (B) at (1,3) {$a$};
	\node (C) at (5,3) {$b$};
	\node (D) at (5,1) {$b$};
	\draw[black, line width = 1pt] (A) -- (2,2);
	\draw[black, line width = 1pt] (2,2) -- (B);
	\draw[black, dashed, line width = 1pt] (2,2) -- (4,2);
	\draw[black, line width = 1pt] (C) -- (4,2);
	\draw[black, line width = 1pt] (D) -- (4,2);
	\end{tikzpicture}\; =\sum_{y}(F_{a,a,b}^{b})_{1}^{y} \;
	\begin{tikzpicture}[baseline = {(current bounding box.center)},scale = 2/3]
	\node (A) at (1,1) {$a$};
	\node (B) at (1,5) {$a$};
	\node (C) at (3,5) {$b$};
	\node (D) at (3,1) {$b$};
	\draw[black, line width = 1pt] (A) -- (2,2);
	\draw[black, line width = 1pt] (2,2) -- (D);
	\draw[black, line width = 1pt] (2,2) -- (2,4);
	\draw[black, line width = 1pt] (C) -- (2,4);
	\draw[black, line width = 1pt] (B) -- (2,4);
	\node[anchor = east] at (2,3) {$y$};
	\end{tikzpicture} \;,
	\end{equation}
	where the dashed line carry the identity object. Because the left side of this equation is an idempotent (it is the identity morphism on $a\otimes b $), so must be the right side which fixes the normalization of the trivalent vertices:
	\begin{equation}
	\begin{tikzpicture}[baseline = {(current bounding box.center)},scale = 2/3]
	\node (A) at (1,4) {$a$};
	\node (B) at (0,2) {$b$};
	\node (C) at (2,2) {$c$};
	\node (D) at (1,0) {$d$};
	\draw[black, line width = 1pt] (A) -- (1,3);
	\draw[black, line width = 1pt] (1,3) -- (B);
	\draw[black, line width = 1pt] (1,3) -- (C);
	\draw[black, line width = 1pt] (B) -- (1,1);
	\draw[black, line width = 1pt] (C) -- (1,1);
	\draw[black, line width = 1pt] (1,1) -- (D);
	\end{tikzpicture} \; \equiv \; \delta_{a,d} \frac{1}{(\mathsf{F}_{c,c,b}^{b})_{1}^{a}} \; \begin{tikzpicture}[baseline = {(current bounding box.center)},scale = 2/3]
	\node (A) at (1,2) {$a$};
	\node (D) at (1,0) {$a$};
	\draw[black, line width = 1pt] (A) -- (D);
	\end{tikzpicture} \;.
	\end{equation}
\subsection{The TL relations} \label{app:TL-relations}
The Temperley-Lieb relations \eqref{TL-relations} can be proven using the relations of appendix \ref{sec:fsymbols}. As an example we derive the third relation:
\begin{align*}
\begin{tikzpicture}[baseline = {(current bounding box.center)},scale = 2/3]
	\draw[black, line width = 1pt] (1,2) -- (3,2);
	\draw[black, line width = 1pt] (1,3) -- (2,4);
	\draw[black, line width = 1pt] (1,5) -- (2,4);
	\draw[dashed, black, line width = 1pt] (2,4) -- (3,4);
	\draw[black, line width = 1pt] (3,4) -- (4,3);
	\draw[black, line width = 1pt] (3,2) -- (4,3);
	\draw[black, line width = 1pt] (3,4) -- (4,5);
	\draw[black, line width = 1pt] (4,5) -- (5,5);
	\draw[black, line width = 1pt] (5,5) -- (6,4);
	\draw[dashed, black, line width = 1pt] (4,3) -- (5,3);
	\draw[black, line width = 1pt] (5,3) -- (6,2);
	\draw[black, line width = 1pt] (5,3) -- (6,4);
	\draw[black, line width = 1pt] (6,2) -- (8,2);
	\draw[dashed, black, line width = 1pt] (6,4) -- (7,4);
	\draw[black, line width = 1pt] (7,4) -- (8,5);
	\draw[black, line width = 1pt] (7,4) -- (8,3);
	\end{tikzpicture} \; & = \sum_{k} (\mathsf{F}_{a,a,a}^{a})_{1}^{k} \;
	\begin{tikzpicture}[baseline = {(current bounding box.center)},scale = 2/3]
	\draw[black, line width = 1pt] (1,2) -- (8,2);
	\draw[black, line width = 1pt] (1,3) -- (2,4);
	\draw[black, line width = 1pt] (1,5) -- (2,4);
	\draw[dashed, black, line width = 1pt] (2,4) -- (3,4);
	\draw[black, line width = 1pt] (3,4) -- (9/2,3);
	\draw[black, line width = 1pt] (6,4) -- (9/2,3);
	\draw[black, line width = 1pt] (3,4) -- (4,5);
	\draw[black, line width = 1pt] (4,5) -- (5,5);
	\draw[black, line width = 1pt] (5,5) -- (6,4);
	\draw[dashed, black, line width = 1pt] (6,4) -- (7,4);
	\draw[black, line width = 1pt] (7,4) -- (8,5);
	\draw[black, line width = 1pt] (7,4) -- (8,3);
	\draw[black, line width = 1pt] (9/2,3) -- (9/2,2);
	\node[anchor = east] at (9/2,5/2) {$k$};
	\end{tikzpicture} \\
	& = \sum_{k,r} (\mathsf{F}_{a,a,a}^{a})_{1}^{k}(\mathsf{F}_{1,a,a}^{k})_{a}^{r} \;
	\begin{tikzpicture}[baseline = {(current bounding box.center)},scale = 2/3]
	\draw[black, line width = 1pt] (1,2) -- (8,2);
	\draw[black, line width = 1pt] (1,3) -- (2,4);
	\draw[black, line width = 1pt] (1,5) -- (2,4);
	\draw[dashed, black, line width = 1pt] (2,4) -- (3,4);
	\draw[black, line width = 1pt] (3,4) -- (4,4);
	\draw[black, line width = 1pt] (4,4) -- (5,5);
	\draw[black, line width = 1pt] (4,4) -- (5,3);
	\draw[black, line width = 1pt] (5,5) -- (6,4);
	\draw[black, line width = 1pt] (5,3) -- (6,4);
	\draw[dashed, black, line width = 1pt] (6,4) -- (7,4);
	\draw[black, line width = 1pt] (7,4) -- (8,5);
	\draw[black, line width = 1pt] (7,4) -- (8,3);
	\draw[black, line width = 1pt] (3,4) -- (9/2,2);
	\node[anchor = east] at (7/2,3) {$k$};
	\node[anchor = south] at (7/2,4) {$r$};
	\node[anchor = south] at (7/2,6) {};
	\end{tikzpicture} \\
	&= \frac{1}{(\mathsf{F}_{a,a,a}^{a})_{1}^{1}}\sum_{k} (\mathsf{F}_{a,a,a}^{a})_{1}^{k}(\mathsf{F}_{1,a,a}^{k})_{a}^{1}
	\begin{tikzpicture}[baseline = {(current bounding box.center)},scale = 2/3]
	\draw[black, line width = 1pt] (1,2) -- (8,2);
	\draw[black, line width = 1pt] (1,3) -- (2,4);
	\draw[black, line width = 1pt] (1,5) -- (2,4);
	\draw[dashed, black, line width = 1pt] (2,4) -- (3,4);
	\draw[dashed, black, line width = 1pt] (3,4) -- (7,4);
	\draw[black, line width = 1pt] (7,4) -- (8,5);
	\draw[black, line width = 1pt] (7,4) -- (8,3);
	\draw[black, line width = 1pt] (3,4) -- (9/2,2);
	\node[anchor = east] at (7/2,3) {$k$};
	\node[anchor = south] at (7/2,6) {};
	\end{tikzpicture} \\
	& = (\mathsf{F}_{1,a,a}^{1})_{a}^{1} \;
	\begin{tikzpicture}[baseline = {(current bounding box.center)},scale = 2/3]
	\draw[black, line width = 1pt] (1,2) -- (8,2);
	\draw[black, line width = 1pt] (1,3) -- (2,4);
	\draw[black, line width = 1pt] (1,5) -- (2,4);
	\draw[dashed, black, line width = 1pt] (2,4) -- (3,4);
	\draw[dashed, black, line width = 1pt] (3,4) -- (7,4);
	\draw[black, line width = 1pt] (7,4) -- (8,5);
	\draw[black, line width = 1pt] (7,4) -- (8,3);
	\node[anchor = south] at (7/2,6) {};
	\end{tikzpicture} \;,
\end{align*}
where the unmarked dashed strands carry the tensor identity $1$, and the unmarked full strands carry the object $a$. The first two lines are obtained through $F$-moves, the third line used the normalisation for trivalent vertices, and the last line uses the trivial fusion rules of the identity.

\subsection{$F$-Symbols}\label{sec:Fsymbols}
We give here the expressions of \cite{Fsymbols} that we have used for the $F$-symbols:
\begin{align}
(F^{d}_{a,b,c})_{e}^{f} =& (-1)^{(a+b+c+d)/2} \Delta(a,b,e)\Delta(c,d,e)\Delta(b,c,f)\Delta(a,d,f)\sqrt{[e+1]_{\q}[f+1]_{\q}} \notag\\
& \sum_{\Lambda}\big( (-1)^{n/2}\big[\frac{n+2}{2}\big]_{\q}! \frac{1}{\big[\frac{a+b+c+d-n}{2}\big]_{\q}!\big[\frac{a+c+e+f-n}{2}\big]_{\q}!\big[\frac{b+d+e+f-n}{2}\big]_{\q}!}\notag \\
& \frac{1}{\big[\frac{n-a-b-e}{2}\big]_{\q}!\big[\frac{n-c-d-e}{2}\big]_{\q}!\big[\frac{n-b-c-f}{2}\big]_{\q}!\big[\frac{n-a-d-f}{2}\big]_{\q}!}
 \big),
\end{align}
where $[x]_{\q} = (-1)^{x-1} \frac{ \q^{x} - \q^{-x}}{\q - \q^{-1}} $, $[x]_{\q}! \equiv [x]_{\q} [x-1]_{\q}$, $[1]_{\q}! \equiv 1$ and
\begin{equation}
	\Delta(a,b,c) \equiv (\frac{\big[\frac{a+b-c}{2}\big]_{\q}\big[\frac{a-b+c}{2}\big]_{\q}\big[\frac{-a+b+c}{2}\big]_{\q}}{\big[\frac{a+b+c+2}{2}\big]_{\q}})^{1/2}.
\end{equation}

\section{Restricted-(not so)-SOS models}
\begin{figure}
\begin{center}
\begin{tikzpicture}[ baseline = {(current bounding box.center)}, scale = 2/3]
	\draw[black, dotted, line width = 1pt]  (3, 3) -- (17,3);
	\foreach \r in {5,7,9,11,13,15}{
		\fill[black] (\r , 5 ) circle (1/8);
		\fill[black] (\r , 1 ) circle (1/8);
		\fill[black] (\r+2 , 3 ) circle (1/8);
		\fill[black] (\r-2 , 3 ) circle (1/8);
		\draw[black, dotted, line width = 1pt]  (\r,5) -- (\r-2,3) -- (\r,1) -- (\r+2,3) -- (\r,5);
	}
	\draw[black, dotted, line width = 1pt] (1,1) -- (3,3);
	\draw[black, dotted, line width = 1pt] (17,3) -- (19,1);
	\fill[black] (1 , 1 ) circle (1/8);
	\fill[black] (3 , 3 ) circle (1/8);
	\fill[black] (19 , 1 ) circle (1/8);
	\draw[black, line width = 2pt] (0,1) -- (0,5);
	\foreach \r in {1,...,5}{
	\filldraw[black] (0,\r) circle (4 pt);
	\node[anchor = east] at (-.25,\r) {\small{\r}};
	};
	\draw[black, line width = 2pt] (1,0) -- (19,0);
	\foreach \r [count = \i ] in {1,3,5,7,9,11,13,15,17,19}{
	\filldraw[black] (\r,0) circle (4 pt);
	\node[anchor = north] at (\r,-.25) {\small{\i}};
	};
\end{tikzpicture}
\end{center}
\caption{The $3$-walks of $9$ steps on $A_{5}$; the dashed lines show all the possible paths which can be taken starting and ending on the element $1$. There are $85$ distinct such paths.}\label{fig:aequals3walk}
\end{figure}
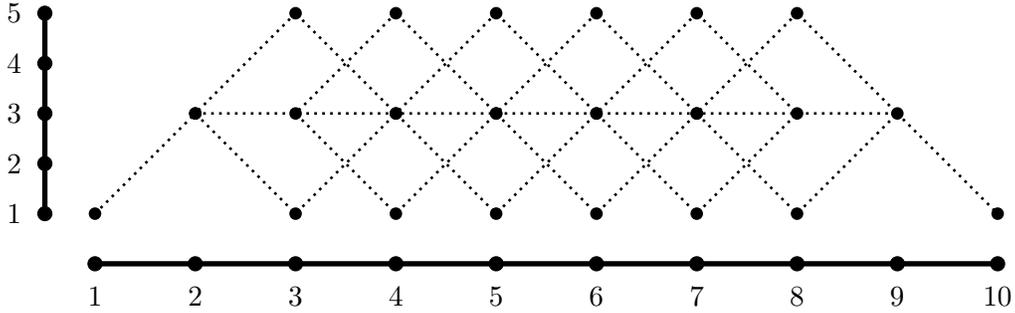

While this work mainly focus on the cases where $a=2$, the other types of homogeneous anyon chains can indeed be interpreted as pseudo-RSOS models. We focus on the case $a =3 $ as it illustrates nicely the main differences between the generic cases, and $a=2$. For simplicity, we shall also restrict ourselves to the open cases, the periodic case being very similar. A $3$-walk of $n$ steps $\lbrace x_{i} \rbrace $ is a tuple of $n+1$ elements of $\mathsf{A}_{p}$ such that for all $i=1, \hdots, n $, $x_{i+1} $ appears in the tensor product $x_{i} \otimes 3$; figure \ref{fig:aequals3walk} illustrates how to interpret these as walks on $\mathsf{A}_{p}$.
One defines an action of the Temperley-Lieb generators on these paths in the following way:
\begin{equation}
	e_{j}|x_{0},x_{1}, \hdots, x_{n} \rangle = \delta_{x_{j-1},x_{j+1}}\sum_{k \in (x_{j-1} \otimes 3 )} \frac{[x_{j}]^{1/2}_{\q}[k]^{1/2}_{\q}}{[x_{j-1}]_{\q}} |x_{0},x_{1},\hdots, x_{j-1},k,x_{j+1},\hdots, x_{n}\rangle,
\end{equation}
where the sum is over the elements of $\mathsf{A}_{p} $ appearing in $x_{j}\otimes 3 $, and $\q = e^{\frac{i \pi}{p+1}} $. One can show that this action defines a module for the Temperley-Lieb algebra with loop weight $\beta = [3]_{\q}$. However, one must remember that the Temperley-Lieb algebra is semi-simple unless the loop weight is of the form $\beta = \Q + \Q^{-1} $ for some root of unity $\Q\neq \pm 1$, so this model is generally semi-simple.

\section{Folded RSOS models}

It is well known that the periodic RSOS models actually contain two copies of the lattice versions of Virasoro minimal models; when the parameter $p$ is even\footnote{This folding can be done for any parity of $p$, but if $p$ is odd then it will not fix this problem of multiplicities.}, this problem can be fixed by \emph{folding} the model. In terms of anyons, one starts by noticing the congruence relation $a \sim p+1-a$, valid in $\mathsf{A}_{p}$ for all $p$. One can then simply take the quotient fusion category and build an anyon model in the usual way:
\begin{equation}
\tilde{\mathcal{H}}_{a + I}[n] \equiv \bigoplus_{x_{0} \in \mathsf{A}_{p}/I}\mathsf{Hom}_{\tilde{\mathcal{C}}_{p}}(x_{0} \otimes (a+I)^{\otimes n}, x_{0}),
\end{equation}
where $I$ is the ideal generated by the congruence relation and $\tilde{\mathcal{C}}_{p}$ is the quotient tensor category. Note that the original anyon model had restrictions on the parity of $n$, whereas the folded one does not.

In terms of RSOS models, this construction is however a bit more subtle. The model we described in section \ref{sec:RSOS} is a module for the affine Temperley-Lieb algebra, so in addition to the generators $e_{i}$, there are other generators $u$ and $u^{-1}$ which \emph{shifts} each path by one step to the right and to the left, respectively, so that
\begin{equation*}
	u e_{i}[n] = e_{i+1}[n]u , \qquad u^{-1}e_{i+1}[n] = e_{i}[n]u^{-1}.
\end{equation*}
 However, these new generators do not appear in the Hamiltonian \eqref{eq:RSOShamiltonian}, so one might consider linear maps which preserves the action of the $e_{i}$s, but not of $u$. The simplest choice simply sends $e_{i} \to e_{i} $, $u\to -u$. Using the notation of section \ref{sec:mixedmerge}, this map is simply
\begin{equation*}
	\mathcal{H}_{x,y}[n] \to \mathcal{H}_{p+1-x, p+1-y}[n].
\end{equation*}
It follows in particular that the Hamiltonian has the same spectrum on $\mathcal{H}_{x,y}[n]$ and $\mathcal{H}_{p+1-x, p+1-y}[n]$. Getting rid of these degeneracies yields the folded Hamiltonian, written in terms of sectors\footnote{The expression of the Hamiltonian in terms of Temperley-Lieb generators is the same, but they act on a different module described in terms of twisted sectors, see section \ref{sec:mixedmerge}.}:
\begin{equation}
	\mathcal{H}_{\rm FRSOS}[n] \sim \left.\begin{cases}
		\bigoplus_{x =1}^{\frac{p}{2}} \mathcal{H}_{x,x}[n] & \text{if } n \text{ is even}\\
		\bigoplus_{x =1}^{\frac{p}{2}} \mathcal{H}_{x,p-x}[n] & \text{if } n \text{ is odd}
	\end{cases}\right\rbrace.
\end{equation}
\subsection{The Fibonacci anyons}\label{sec:fibon}
Perhaps the simplest case of folded RSOS model is the one corresponding to the Fibonacci anyons: these contain two anyons, labelled $1$ and $\tau$, with fusion rules
\begin{equation}
	\tau \otimes \tau = 1 + \tau, \qquad 1\otimes \tau = \tau \otimes 1 = \tau, \qquad 1\otimes 1 =1.
\end{equation}
These can be obtained by folding the $\mathsf{A}_{4}$ anyons:
\begin{equation}
	1 \equiv 1 \sim  4, \qquad \tau \equiv 2 \sim 3.
\end{equation}
We now give expressions for the various Temperley-Lieb generators appearing in this model ($\q = e^{\frac{i \pi}{5}}, \phi =\frac{1}{2}( 1+\sqrt{5}) $), with the basis:
\begin{align}
\lbrace\;& |1,\tau,\tau,1 \rangle,\; |1,\tau,1, \tau\rangle, \; |1,\tau,\tau,\tau \rangle , \; |\tau,1,\tau,1 \rangle, \;\notag\\& |\tau, \tau, \tau, 1 \rangle, \; |\tau, 1, \tau, \tau \rangle, \; |\tau, \tau, 1, \tau \rangle, \; | \tau, \tau, \tau, \tau \rangle\; \rbrace
\end{align}
\begin{equation}
e_{1} = \left(0\right) \oplus
\left(
\begin{array}{cc}
 \phi & 0 \\
0 & 0
\end{array}
\right) \oplus
\left(
\begin{array}{cc}
 \frac{1}{\phi } & \frac{-1}{\sqrt{\phi }} \\
 \frac{-1}{\sqrt{\phi }} & 1
\end{array}
\right) \oplus
\left(
\begin{array}{ccc}
 \frac{1}{\phi } & 0 & \frac{-1}{\sqrt{\phi }} \\
0 & 0& 0 \\
 \frac{-1}{\sqrt{\phi }} & 0 & 1
\end{array}
\right)
\end{equation}
\begin{equation}
e_{2} = \left(0\right) \oplus
\left(
\begin{array}{cc}
 \frac{1}{\phi} & \frac{-1}{\sqrt{\phi}} \\
\frac{-1}{\sqrt{\phi}} & 1
\end{array}
\right) \oplus
\left(
\begin{array}{cc}
 \phi & 0\\
0 & 0
\end{array}
\right) \oplus
\left(
\begin{array}{ccc}
0 & 0 &0 \\
0 &  \frac{1}{\phi } & \frac{-1}{\sqrt{\phi }} \\
0 & \frac{-1}{ \phi ^{1/2}} & 1
\end{array}
\right)
\end{equation}
While the one carrying defects of type $r$, of width one, in the direct channel, is:
\begin{align}
&\tilde{e}(w)- f(w) \mathbb{I}_{8}   = \left(0\right) \oplus
\left(
\begin{array}{cc}
 \frac{2  + \phi (1+ g_{-}(w) + g_{+}(w))}{\phi} & -\frac{1+ \phi g_{-}(w)}{\phi^{1/2}} \\
 -\frac{1+ \phi g_{+}(w)}{\phi^{1/2}}& 1
\end{array}
\right)\notag \\& \oplus
\left(
\begin{array}{cc}
 \frac{2  + \phi (1+ g_{-}(w) + g_{+}(w))}{\phi} & -\frac{1+ \phi g_{+}(w)}{\phi^{1/2}} \\
 -\frac{1+ \phi g_{-}(w)}{\phi^{1/2}}& 1
\end{array}
\right)   \oplus
\left(
\begin{array}{ccc}
 \frac{1}{\phi } & \frac{g_{-}(w)}{\phi } & -\frac{1+ g_{-}(w)}{\phi ^{1/2}} \\
\frac{g_{+}(w)}{\phi }  & \frac{1}{\phi } & -\frac{1+ g_{+}(w)}{\phi ^{1/2}}\\
 -\frac{1+ g_{+}(w)}{\phi ^{1/2}} & -\frac{1+ g_{-}(w)}{\phi ^{1/2}} & {\scriptstyle{2 + g_{-}(w) + g_{+}(w)}}
\end{array}
\right)
\end{align}
with $g_{\pm}(w) \equiv \frac{w^{-1}-w}{\q^{\pm}w - \q^{\mp}w^{-1}}$, $f(w) = \frac{ \q^{-2} - \q^{2}}{(w + w^{-1})(\q w - \q^{-1}w^{-1} )}$. The generators carrying a defect of type $s$ are then:
\begin{equation}
\lim_{w \to 0} \tilde{e}(w) = \left(0\right) \oplus
\left(
\begin{array}{cc}
 \frac{1}{\phi } & \frac{\q^{-2}}{\sqrt{\phi }} \\
 \frac{\q^{2}}{\sqrt{\phi }} & 1
\end{array}
\right) \oplus
\left(
\begin{array}{cc}
 \frac{1}{\phi } & \frac{\q^2}{\sqrt{\phi }} \\
 \frac{\q^{-2}}{\sqrt{\phi }} & 1
\end{array}
\right) \oplus
\left(
\begin{array}{ccc}
 \frac{1}{\phi } & -\frac{\q^{-1}}{\phi } & -\frac{\q^2}{\phi ^{3/2}} \\
 -\frac{\q}{\phi } & \frac{1}{\phi } & -\frac{\q^{-2}}{ \phi ^{3/2}} \\
 -\frac{\q^{-2}}{ \phi ^{3/2}} & -\frac{\q^2}{\phi ^{3/2}} & \frac{1}{\phi ^2}
\end{array}
\right)
\end{equation}
\begin{equation}
\lim_{w \to \infty} \tilde{e}(w) = \left(0\right) \oplus
\left(
\begin{array}{cc}
 \frac{1}{\phi } & \frac{\q^2}{\sqrt{\phi }} \\
 \frac{\q^{-2}}{\sqrt{\phi }} & 1
\end{array}
\right) \oplus
\left(
\begin{array}{cc}
 \frac{1}{\phi } & \frac{\q^{-2}}{\sqrt{\phi }} \\
 \frac{\q^{2}}{\sqrt{\phi }} & 1
\end{array}
\right) \oplus
\left(
\begin{array}{ccc}
 \frac{1}{\phi } & -\frac{\q}{\phi } & -\frac{\q^{-2}}{\phi ^{3/2}} \\
 -\frac{\q^{-1}}{\phi } & \frac{1}{\phi } & -\frac{\q^{2}}{ \phi ^{3/2}} \\
 -\frac{\q^{2}}{ \phi ^{3/2}} & -\frac{\q^{-2}}{\phi ^{3/2}} & \frac{1}{\phi ^2}
\end{array}
\right)
\end{equation}
Note that because $\phi$ is real and $\q$ is a root of unity, these generators are still Hermitian.
\end{document}